\documentclass[fleqn,usenatbib]{mnras}

\usepackage{newtxtext,newtxmath}
\usepackage{soul}
\usepackage[T1]{fontenc}
\DeclareRobustCommand{\VAN}[3]{#2}
\let\VANthebibliography\thebibliography
\def\thebibliography{\DeclareRobustCommand{\VAN}[3]{##3}\VANthebibliography}
\usepackage[dvipsnames]{xcolor}

\usepackage{graphicx}
\usepackage{amsmath}
\usepackage{amssymb}
\usepackage{changes}
\usepackage{hyperref}

\usepackage{adjustbox}
\usepackage{anyfontsize}
\usepackage{longtable}
\usepackage{afterpage}
\usepackage{orcidlink}

\let\oldAA\AA
\renewcommand{\AA}{\text{\normalfont\oldAA}}
\title[Counting and Ranking Hosts of PTA Sources]{Prospects of resolving and localising individual supermassive black hole binaries with pulsar timing arrays: the host ranking challenge}

\author[N. Veronesi et al.]{
Niccolò Veronesi\orcidlink{0000-0001-7678-8218}$^{1}$,
\thanks{E-mail: niccolo.veronesi@wsu.edu}
Maria Charisi\orcidlink{0000-0003-3579-2522}$^{1,2}$,
Polina Petrov\orcidlink{0000-0001-5681-4319}$^{3}$,
Stephen R. Taylor\orcidlink{0000-0001-8217-1599}$^{3}$,
Jessie Runnoe\orcidlink{0000-0001-8557-2822}$^{3,4}$,
\newauthor
Daniel J. D'Orazio\orcidlink{0000-0002-1271-6247}$^{5,6,7}$,
Jacob Pilawa\orcidlink{0000-0001-7040-9117}$^{8,9}$,
Chung-Pei Ma\orcidlink{0000-0002-4430-102X}$^{8,9}$,
\\
$^{1}$ Department of Physics and Astronomy, Washington
State University, 1245 Webster Hall, Pullman, WA 99164, USA\\
$^{2}$ Institute of Astrophysics, FORTH, GR-71110,
Heraklion, Greece\\
$^{3}$ Department of Physics and Astronomy, Vanderbilt
University, 2301  Vanderbilt Place, Nashville, TN 37235, USA\\
$^{4}$ Fisk University, Department of Life and Physical
Sciences, 1000 17th Avenue N, Nashville, TN 37208, USA\\
$^{5}$ Space Telescope Science Institute, 3700 San
Martin Dr., Baltimore, MD 21218, USA\\
$^{6}$ Department of Physics and Astronomy, Johns
Hopkins University, 3400 North Charles Street, Baltimore,
Maryland 21218, USA\\
$^{7}$, Niels Bohr International Academy, Niels Bohr
Institute, Blegdamsvej 17, 2100 Copenhagen, Denmark\\
$^{8}$, Department of Astronomy, University of California,
Berkeley, 501 Campbell Hall \#3411, Berkeley, CA 94720, USA\\
$^{9}$Department of Physics, University of California,
Berkeley, CA 94720, USA}

\date{Accepted XXX. Received YYY; in original form ZZZ}

\pubyear{XXXX}

\begin{document}
\label{firstpage}
\pagerange{\pageref{firstpage}--\pageref{lastpage}}
\maketitle

%------------------------------------------------------------------
\begin{abstract}
Pulsar Timing Arrays (PTAs) are soon expected to detect individually
resolved supermassive black hole (SMBH) binaries, opening the
possibility for multi-messenger discoveries. The biggest challenge
will be to pinpoint the host galaxy in a large localisation area.
We simulate realistic binary populations consistent with the
gravitational wave (GW) background, projecting the PTA sensitivity
for the next 0-10 years. We inject the loudest binary on top of the
background and use one of the standard detection pipelines to constrain
its properties. We cross-match the localisation areas with comprehensive
all-sky galaxy catalogues and estimate the number of candidate hosts in
the localisation area assessing, for the first time, the number of missing
galaxies due to incomplete coverage. We develop a ranking system that
excludes galaxies with properties inconsistent with the GW posteriors,
and prioritizes the remaining galaxies for follow-up observations. We
find a $\approx$21, $\approx$38 and $\approx$51 percent probability of
resolving a binary in the next 0, 5 and 10 years, respectively, reduced
to 0.3, 3.8 and 14.1 percent if we require potentially well-constrained
localisation areas. The localisation areas span hundreds of square degrees,
but shrink significantly with the addition of more data. They contain on
average $\approx$190,000 early type galaxies and $\approx$40,000 active
galactic nuclei, with $\approx$25,000 missing candidate hosts. Our ranking
method can exclude about half of the potential hosts and efficiently rank
those remaining when the galaxy catalogue provides SMBH masses and redshifts,
but becomes more inefficient when we rely on apparent magnitudes.
\end{abstract}


%------------------------------------------------------------------

\section{Introduction}
\label{sec:intro}

Pulsar timing arrays (PTAs) have recently found evidence for a stochastic
background of gravitational waves (GWs) at nanohertz frequencies
\citep{Agazie23a,EPTA23, Reardon23, Xu23, Miles25}, the properties of
which are compatible among different PTA datasets \citep{Agazie24}. This
signal is likely emitted by an unresolved population of supermassive black
hole (SMBH) binaries, which are expected to form following the merger of
their parent galaxies. Even though the main steps of binary evolution have
been known for decades \citep{Begelman80}, the timescales over which
binaries form and the processes that drive them to their final coalescence
are still unconstrained \citep{Sesana15,Goicovic17}.

Thanks to the addition of new monitored pulsars and increasing timing
baselines, the sensitivity of PTAs is steadily improving. This not only
enables a more precise characterization of the stochastic background
(which will improve our understanding of binary formation and evolution),
but may also eventually lead to the resolution of the loudest binaries
\citep{Rosado15,Mingarelli17,Kelley18,Becsy22b,Gardiner24}. The isolation
of GW signals on top of the background and their localisation on the sky
will trigger searches for their electromagnetic (EM) counterpart and the
identification of the host galaxy. This, in turn, will enable multi-messenger
analyses providing the most complete information on the physics of the
system \citep{Kelley19,Charisi26}. The complementarity of the information
carried by GWs and EM radiation will provide unprecedented insights onto
topics such as the evolution of SMBH binaries in post-merger galaxies,
and their dynamical interactions with their environment \citep{Laal25}.

The main challenge in identifying the host galaxy arises from the
limited localisation capability of PTAs. The size of the 90 percent
credibility level localisation area (A90) of the first individually
resolved SMBH binary is expected to be several hundreds of square
degrees \citep{Sesana10,Goldstein18,Goldstein19,Petrov24,Truant25}. It
will thus contain a large number of potential hosts.
However, the first resolvable source will likely have a large total mass
($M_{\rm tot}\gtrsim10^9M_\odot$) and will reside in a luminous massive
galaxy at relatively low redshift ($z\leq0.5$), at least if it is detected
in the near-future (5-10 years). This significantly limits the number of
candidate hosts \citep{Goldstein19,Petrov24}, mitigating the probability
of a false-positive association and enhancing our chances of finding the
galaxy in which the PTA source resides.

Another encouraging finding from previous studies is that the host of the
first PTA detectable binary should be bright enough to be included in
current all-sky EM surveys \citep{Goldstein19, Veronesi25,Truant26}.
Therefore, the host is likely already present in an existing EM catalogue,
as long as it is not situated in regions typically avoided due to high
contamination from stars or dust, like the Galactic plane or the Magellanic
Clouds. Here we quantify the impact of incomplete EM coverage on the host
galaxy search.

Finally, when it comes to EM observables, \citet{Veronesi25} demonstrated
that one can use empirical scaling relations
\citep[e.g.,][]{Ferrarese00,McConnell13} to convert the posterior
distribution of GW parameters (such as the chirp mass of the binary)
into distributions of the apparent magnitudes of the host galaxy in
several EM bands. Here we develop a method that leverages the above
distributions to rank the candidates based on their likelihood of containing
the GW-detected source. As we demonstrate in this work, the resulting ranked
list of potential hosts can be used in subsequent searches for promising
binary signatures (see, e.g., \citealt{DOrazio23} for a review) to prioritize
specific targets in follow-up studies likely employing both archival data
and new observations.

In this work, we simulate the projected sensitivity of future PTA datasets
along with binary populations consistent with the GW background to predict
the properties of the first resolved source. We select binaries that could
be localised in the next 5-10 years, but would have not show any evidence
for their presence in the 15-year Data Set of the North American Nanohertz
Observatory for Gravitational Waves (NANOGrav). We then estimate the size
of A90 of the injected binaries using a standard PTA pipeline. 

Next, we cross-match the GW maps with two all-sky galaxy catalogues containing:
(1) quiescent early-type galaxies (ETGs), and (2) active galactic nuclei
(AGN). These state-of-the-art catalogues are expected to be complete within
the projected sensitivity volume of PTAs, especially for the expected massive
hosts of the first resolved binaries. This allows us to obtain a realistic
estimate of the number of candidate hosts within the localisation area, and,
for the first time, quantify the limitations that come from the incomplete
coverage of galaxy catalogues in sky regions with high contamination.
Similarly to what has been first done in \citet{Goldstein19}, we construct
a ranking system which sorts the candidate hosts, taking into account how well
their positions and photometric properties match the posterior distributions
constrained starting from standard PTA detection pipelines. We demonstrate
that this method is capable of excluding galaxies that are found to be
inconsistent with the PTA signal, using solely their photometric and astrometric properties.

In Section \ref{sec:methodology}, we describe our methodology. In
particular, we present the PTA datasets we simulate and analyze,
the EM catalogues we use in cross-matches with the localisation
areas, and the host ranking system. In Section \ref{sec:res}, we
present our results, and in Section \ref{sec:discus} we discuss their
implications for multi-messenger investigations, caveats of our method,
and plans on how to address those in future works. Finally, in Section
\ref{sec:concl}, we present our summary and conclusions.

Whenever necessary, we convert redshift to luminosity distance from
Earth assuming the cosmological parameters of the observations by
Planck \citep{Planck20}: $H_0=67.32 {\rm\ km\ s}^{-1}{\rm Mpc}^{-1}$,
$\Omega_mh^2=0.14314$, and $n_s=0.96605$.

%------------------------------------------------------------------
%------------------------------------------------------------------
%------------------------------------------------------------------

\section{Methodology}
\label{sec:methodology}
In this section, we describe the simulations of GW signals from binaries
that can be resolved on top of the GW background and the data analysis
pipeline that allows us to localise these sources. We describe the main
properties of two all-sky EM catalogues (one containing only quiescent
ETGs and one with only AGN) which we use to cross-match with the GW
localisation areas. These have a high level of completeness up to
$z\approx 0.5$, and likely include all the potential hosts of the PTA
sources, with the exception of areas not covered due to stellar or dust
contamination. We explicitly calculate the limitations that come from
the incomplete sky coverage. We also present a ranking method that combines
the galaxy sky location and photometric properties and sorts the candidate
hosts based on which best matches our expectations with respect to the
posteriors obtained from the PTA detection pipeline.

\subsection{Future PTA configurations}
\label{sec:ptaconfigurations}

To simulate the detection of SMBH binaries, we first construct PTAs that
resemble our expectations for upcoming datasets from the International
Pulsar Timing Array (IPTA). Similarly to our previous work in
\citet{Veronesi25}, we construct a first array called \emph{IPTA}\_20,
denoting a total baseline of 20 years. This includes a total of 116
pulsars, 68 of which are from the NANOGrav 15-year Data Set (NG15;
\citealt{Agazie23c})\footnote{Only 67 pulsars were included in the search
for the GW background, since the monitoring baseline of one pulsar was
not consistent with the minimum requirement of at least 3 years of data.},
31 from the first data release of the MeerKAT Pulsar Timing Array (MPTA;
\citealt{Miles23}), 14 from the first data release of the Parkes Pulsar
Timing Array (PPTA; \citealt{Zic23}), and 3 from the DRnew2+ data release
of the European Pulsar Timing Array (EPTA) and the Indian Pulsar Timing
Array (InPTA; \citealt{EPTA23}). This configuration is similar to the
upcoming third data release of the IPTA.

We also construct a PTA, with a total baseline of 25 years and 42 new
pulsars added with respect to \emph{IPTA}\_20, which we will refer to
as \emph{IPTA}\_25. The new pulsars are added at a rate of 7 pulsars per
year, and are included in the dataset once their monitoring baseline
reaches at least 3 years. The sky positions of the added pulsars are
drawn from a kernel density estimation obtained from the positions of
the 116 observed pulsars used in \emph{IPTA}\_20. \emph{IPTA}\_25 is
our fiducial PTA configuration, which we use in most analyses. Finally,
we explore how our results evolve in an extended PTA dataset, labeled as
\emph{IPTA}\_30. This is constructed by adding 42 more pulsars (reaching
a total of 200), with timing baselines extended by 5 more years, and a
total baseline of 30 years. The positions of the new pulsars are
extracted from the same kernel density estimation as in \emph{IPTA}\_25.

We refer the reader to Section 2.1 of \citet{Veronesi25} for a more
detailed description of the simulated PTAs for all the configurations
presented above.

\subsection{Simulations of SMBH binary populations}
\label{sec:popcreat}
We simulate realistic binary populations to predict the expected properties
of the first resolved PTA source that could be detected in the near future
(i.e. in \emph{IPTA}\_20, \emph{IPTA}\_25 or \emph{IPTA}\_30). We generate
1,000 realizations of SMBH binary populations, which are consistent with
the stochastic GW background measured in NG15. These populations are built
analogously to \citet{Agazie23b}, using semi-analytic models from the
publicly available software package \texttt{holodeck}
\footnote{https://github.com/nanograv/holodeck}. Below we briefly
summarize the main components of these models, but we refer the reader
to \citet{Agazie23b} for a complete explanation of the adopted analytical
expressions, relevant parameters and their fixed values.

We first model the galaxy stellar mass function using a Schechter profile,
which provides the number density of galaxies as a function of their
stellar mass and redshift. Next, we model the galaxy pair fraction and
the galaxy merger time as functions of the galaxy stellar mass, redshift,
and stellar mass ratio of the pair. These three functions combined provide
the number density of galaxy mergers.

We then proceed to populate these merging systems with SMBHs, the masses
of which are correlated with different properties of the parent galaxies
\citep{Satopolito25}. For this, we use the scaling relation between the
SMBH mass, $M_{\rm SMBH}$, and the bulge mass of its host, $M_{\rm b}$:
\begin{equation}
    \label{eq:mmbulge}
    \log_{10}\left(\frac{M_{\rm SMBH}}{M_\odot}\right)=\mu+
    \alpha_\mu\log_{10}\left(\frac{M_{\rm b}}{10^{11}M_\odot}\right)+
    \mathcal{N}(0,\epsilon_\mu),
\end{equation}
where $\mu$ and $\alpha_\mu$ are the intercept and slope of this
relation, and $\mathcal{N}(0,\epsilon_\mu)$ indicates an intrinsic
scatter given by a Gaussian with null mean and a standard deviation
of $\epsilon_\mu$. Following the assumption made in \citet{Agazie23b},
we set the bulge stellar mass to 61.5 percent of the total stellar mass
of the galaxy, based on empirical bulge mass measurements presented in
\citet{Bluck14} and in \citet{Lang14}. This value takes into account
that, in the entire galaxy population, there are both elliptical
galaxies entirely composed of their bulge, and spiral galaxies, in which
only a fraction of the stellar mass is contained in the bulge.

The above steps allow us to calculate the number density of SMBH pairs
as a function of their total mass, mass ratio, and initial redshift.
In practice, \texttt{holodeck} estimates the number density of pairs in
a grid of these parameters. More specifically, the actual number of
binaries in each cell of this three-dimensional (3D) parameter space
is drawn from a Poissonian distribution around the expectation value
calculated by integrating the differential number density over the cell
volume. In this work, we consider a 3D grid with the following ranges
and resolution:
\begin{itemize}
    \item Total mass, $M_{\rm tot} [M_{\odot}]$: $\left[10^4-10^{12}\right]$,
    720 bins;
    \item Mass ratio, q: $\left[10^{-3} -1\right]$, 160 bins;
    \item Initial redshift, $z_{in}$: $\left[10^{-3}-10\right]$, 100 bins.
\end{itemize}
While we use the ranges set as default in \texttt{holodeck} and used in
the analysis of \citet{Agazie23b}, we increase the number of bins in total
mass by a factor of 8 and in redshift by a factor of 2. This leads to a
finer overall resolution of the sampled parameter space, while remaining
computationally manageable (thanks to the significantly smaller number of
binary populations required in this work compared to \citealt{Agazie23b}).

Finally, we evolve the SMBH binaries from very large separations to the
PTA regime, using an empirical function to calculate how fast their
separation $a$ decreases as a function of time. This hardening rate is
modeled using the following phenomenological double power law:
\begin{equation}
    \label{eq:hardening}
    \frac{da}{dt}=H_a \cdot\left(\frac{a}{a_c}\right)^{1-\nu_{\rm inner}}
    \cdot\left(1+\frac{a}{a_c}\right)^{\nu_{\rm inner}-\nu_{\rm outer}},
\end{equation}
where the normalization $H_a$ is calculated from the total binary
lifetime, and the critical separation $a_c$ denotes the boundary between
the two orbital decay regimes of the model. At large separations, the
shrinking is governed by the exponent $\nu_{\rm outer}$ and encapsulates
the evolution of the pair due the physical interactions with the surrounding
environment, such as dynamical friction and three-body interactions with
the stars. The exponent $\nu_{\rm inner}$ governs the regime of small
separations, where the binary evolves primarily due to the emission of GWs
and interactions with the surrounding gaseous environment. Each binary is
then evolved until it reaches the PTA regime at a certain final redshift
$z$, which we use to calculate the luminosity distance from Earth in the
observer's reference frame. From this point on, when we refer to redshift
we mean this quantity, regardless of the initial redshift at which the
galaxy merger happened.

The above models have six free parameters: the normalization of the galaxy
stellar-mass function and its characteristic mass, the intercept $\mu$ of
the scaling relation between bulge mass and SMBH mass and its intrinsic
scatter $\epsilon_\mu$, the total binary lifetime, and the hardening
power-law index that governs the evolution at small separations
$\nu_{\rm inner}$. Each population is created by fixing the other
parameters to the fiducial values assumed in \citet{Agazie23b}, and by
drawing one sample from the 6-dimensional posterior distribution of
the same paper, using the models with priors informed by observational
measurements (i.e. the \texttt{Phenom+Astro} model, see the column
corresponding to the Astrophysical Priors in Table B1 and the respective
posteriors in Figure 9 of \citealt{Agazie23b}). The simulated SMBH binary
populations are therefore consistent with the GW background, as measured
in NG15.

Similarly to \citet{Agazie23b}, in our simulations we use 14 frequency
bins (between 0.99 nHz and 28.71 nHz, each with a width of 1.98 nHz).
The frequency grid is determined by the total baseline of NG15
($T_{\rm data}=16.03$ yr, which corresponds to the time between the first
and the last measured TOA of all monitored pulsars), with the upper limit
of the $i$-th bin defined as $i/T_{\rm data}$. Each generated binary
population is evolved to emit GWs in this frequency range. We note that
to have full compatibility with previous works, such as \citet{Agazie23b},
we adopt the same frequency range of NG15, even if real searches with the
baselines we employ will be able to detect binaries below it. For this reason,
the numbers of detectable and localisable binaries presented in this work
are to be considered lower limits.

\subsection{Selection of first resolvable PTA binaries}
\label{sec:gwselec}

Having generated 1,000 binary populations, we then select the loudest
source of each population, i.e. the one that is more likely to be
resolved first. For this, we first select the binary with the largest
strain in each frequency bin. We examine each of these 14 binaries
individually to determine which one has the highest signal to noise
ratio (S/N) overall in the PTA configuration under consideration and
will be injected as a resolved source in our simulations (see below).

After removing the loudest source from each frequency bin, we calculate
the total strain of the rest of the population as the square root of the
sum of the square of the strains of the remaining binaries. We do this
to establish the level of the background that we will inject in our
simulations. In this step, we calculate the strain of the stochastic
GW background, $h_{\rm bg}$, in the first five frequency bins. This is
similar to \citet{Agazie23b}, in which the population synthesis models
were derived using 14 frequency bins, but the fit to the PTA data was
done focusing only on the first five. This is because the GW background
is fairly unconstrained at higher frequencies in NG15.

We estimate the best-fit background amplitude, $A_{\rm GWB}$, and
spectral index, $\alpha_{\rm GWB}$, for each realization by performing
a least-squares fit using a single power law
(\(h_{\rm bg}=A_{\rm GWB}\cdot f^{\alpha_{\rm GWB}}\)).
We emphasize that the distribution of the background parameters we
obtain from these fits ($A_{\rm GWB}=-14.75\pm0.19$, and
$\gamma_{\rm GWB}=3-2\cdot\alpha_{\rm GWB}=4.66\pm0.24$)\footnote{We
report the value of $\gamma_{\rm GWB}$ to allow a direct comparison
with the results of \citealt{Agazie23b}.} are compatible both with
those derived in the phenomenological models that we also employ here
and with those constrained by the PTA data using the Hellings-Downs
model (see Figure 7 of \citealt{Agazie23b}). Therefore, the removal of
the binary with the highest strain in each frequency bin does not
introduce a significant bias in the simulated GW background. This
likely means that the GW background is not dominated by the loudest
sources in each bin \citep{Agazie25}.
The values obtained in this step are used in the rest of the
analysis to generate the TOAs for the GW background and estimate
the S/N of the resolved binaries we inject on top of this stochastic
signal.

Next, we calculate the S/N of the binaries with the highest strain
for each frequency bin to select the one with the overall highest S/N
in the population. For this, we generate the TOAs of pulsar signals
for a realization of the GW background with the parameters from the fit
above and then inject a continuous wave source with parameters of the
respective binary, as obtained from \texttt{holodeck}. For each individual
binary, we have the total mass, mass ratio, redshift, and GW frequency,
but for the TOAs generation, we also need the sky location, inclination
of the angular momentum vector with respect to our line of sight, GW
initial phase, and GW polarization angle.
We assume that the binaries are located in the most sensitive location
of the \emph{IPTA}\_25 configuration (since this is our fiducial PTA).
We note that the specific sky location is not important for the selection
of the loudest binary, since at this stage we are interested in finding
the highest S/N among the ones of the 14 binaries we examine for each
population. The most sensitive sky location is obtained by following
the process described in \citet{Veronesi25}, practically injecting a
test binary with fixed parameters in the centres of all the pixels of
a Hierarchical Equal Area isoLatitude Pixelisation (Healpix) projection
with NSide = 4 and determining their S/N.

Once the loudest binary for each population is selected, we inject it
at a random sky location and re-calculate its S/N. To each binary we
assign a random value of phase, polarization angle, and inclination. We
consider a binary to be detectable if it reaches S/N$=4$ \footnote{The
choice of this S/N threshold is slightly arbitrary, since, in reality,
detection statistics do not rely on a simple S/N measurement, but thresholds
in the range of 3$\leq$S/N$\leq$5 are typical in similar studies (e.g.,
\citealt{Truant25}). Assuming a different threshold would not significantly
alter our conclusions.}. However, in this work we are mainly interested in
sources that have a S/N high enough not only to be confidently detected
above the background, but also to be localised in the sky with enough
precision to allow EM searches for the host galaxy. For this purpose, we
select binaries with S/N$\geq8$, since this is roughly the threshold that
typically leads to well defined posteriors on the binary sky location
\citep{Sesana10,Taylor16,Petrov24}. We also take into account that, in the
realization of the Universe we live in, we have not found any evidence for
an individual SMBH binary in the existing PTA datasets. Therefore, we require
that these binaries also have an $S/N<4$ in NG15.

\begin{figure*}
    \centering
    \includegraphics[trim= 0 100 0 100
    ,clip,width=0.92\textwidth]{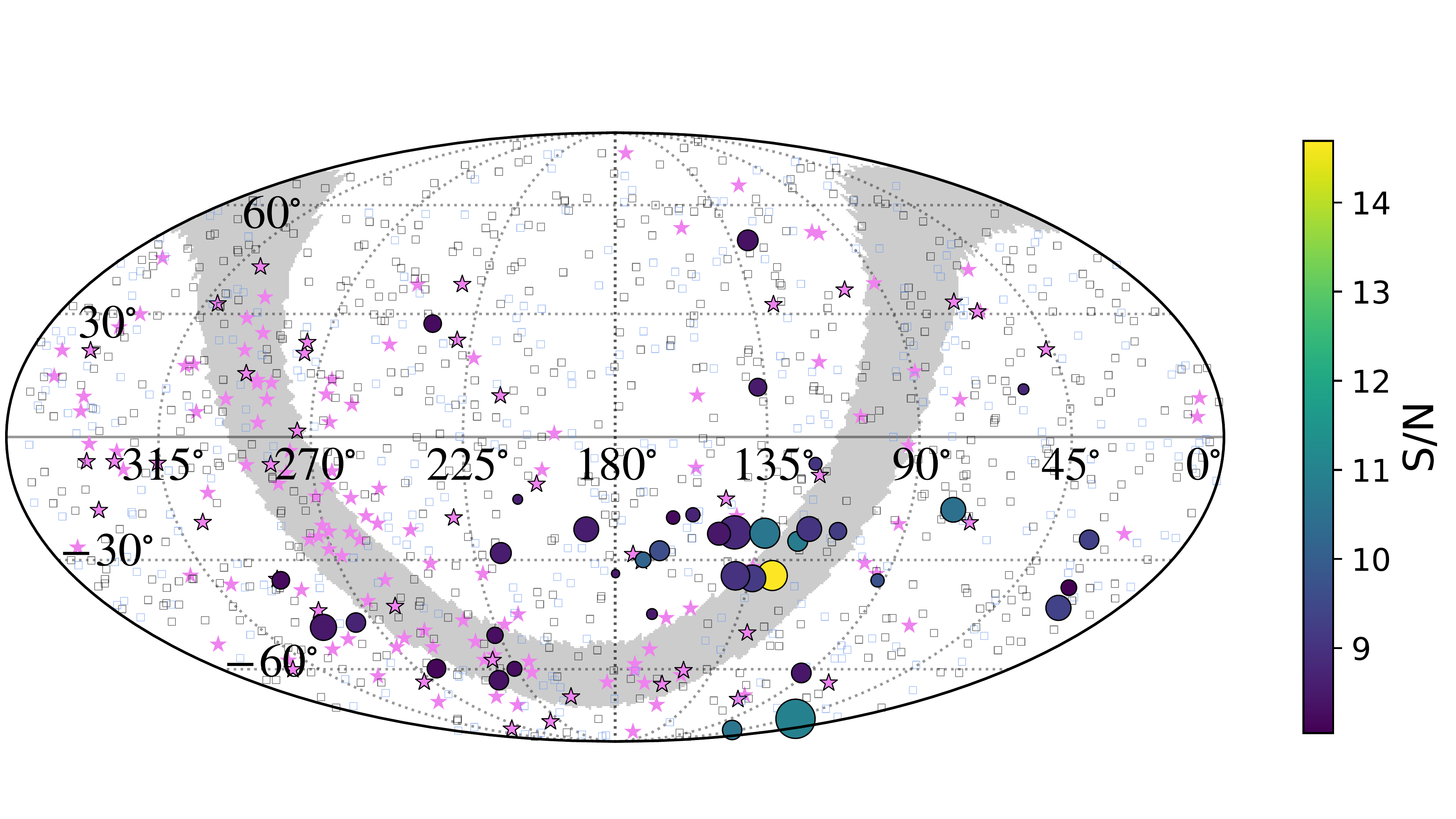}
    \caption{Mollweide projection of the positions of binaries
    (round markers) that, while not detectable by NG15 ($S/N<4$),
    are potentially localisable by \emph{IPTA}\_25 ($S/N\geq8$),
    color-coded according to their S/N in \emph{IPTA}\_25. The
    size of the markers is proportional to the total binary mass.
    The squares mark the position of the loudest binary in each
    of the 1,000 populations we have created, with the blue ones
    indicating the sources among them that, while not detectable
    in NG15, would be detectable, but not potentially localisable
    ($4\leq {\rm S/N}<8$), in \emph{IPTA}\_25. The gray shaded
    area marks the Galactic plane, i.e. the region with Galactic
    latitude $|b|\leq10^\circ$. The pink stars with no outline
    show the pulsars in \emph{IPTA}\_20, while the ones with black
    outlines indicate the 42 pulsars that are added in \emph{IPTA}\_25.}
    \label{fig:binposits}
\end{figure*}

\begin{figure*}
    \centering
    \includegraphics[trim= 60 0 60 0 ,clip,
    width=0.92\textwidth]{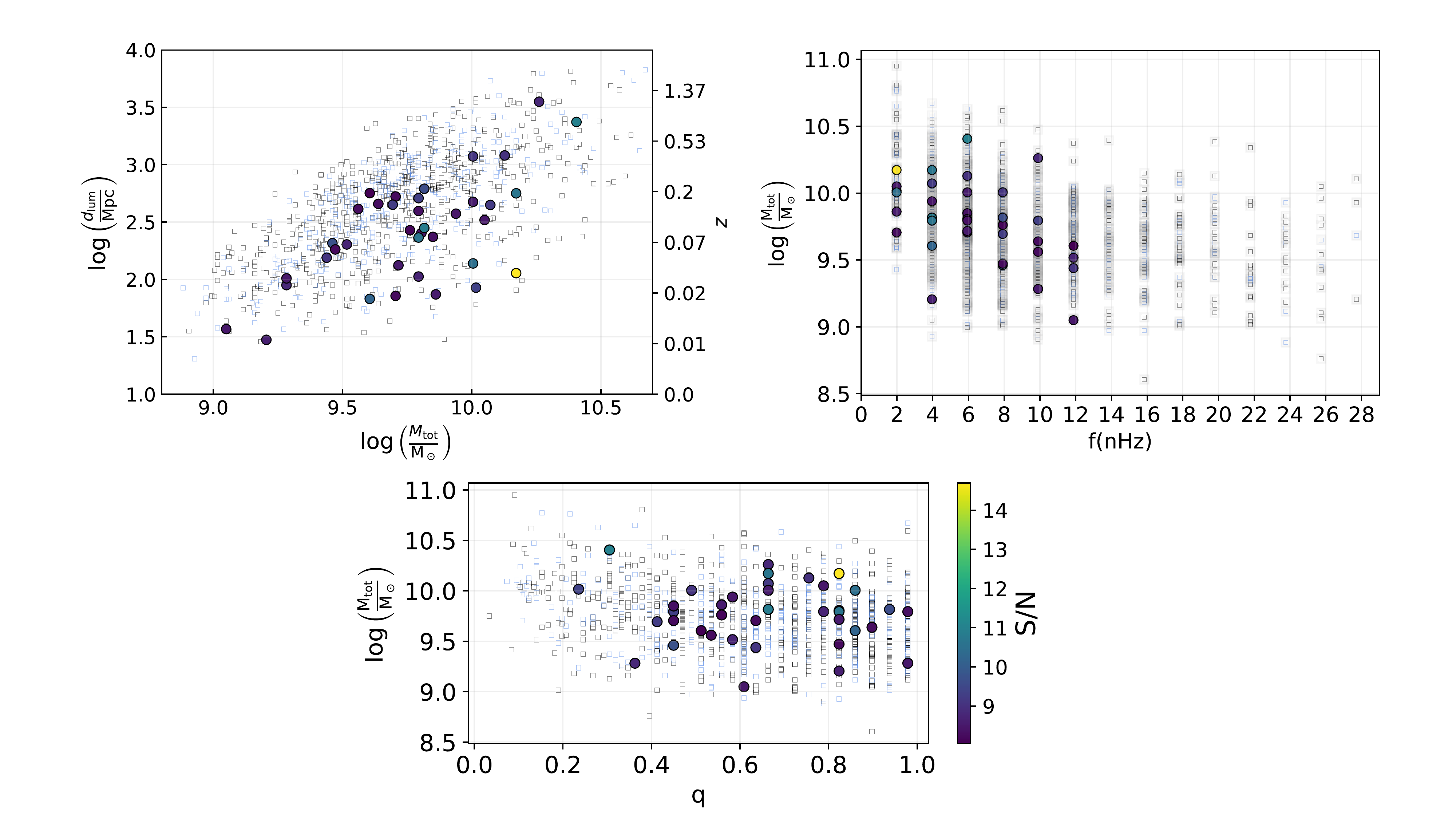}
    \caption{Properties of the binaries that are not detectable
    in NG15, but are potentially localisable in \emph{IPTA}\_25
    (round markers), color-coded by their S/N. The shape and the
    color-code of all markers are the same as in Figure
    \ref{fig:binposits}. The upper left panel shows the luminosity
    distance from Earth (and redshift) of the source as a function
    of its total mass, the upper right panel shows the total mass
    as a function of the GW frequency, and the lower panel shows
    the total mass as a function of the mass ratio.}
    \label{fig:binprops}
\end{figure*}

The sky positions and the S/N of the selected sources are shown in Figure
\ref{fig:binposits}. The marker color represents the S/N of the binary in
\emph{IPTA}\_25, while its size is proportional to the total mass of the
system. The gray area delineates the Galactic plane, i.e. regions with a
Galactic latitude $|b|\leq10^\circ$. We show the positions of the sources
detectable by \emph{IPTA}\_25 but not by NG15 with blue squares. The
positions of the loudest binaries of the remaining populations are marked
by gray squares. The pink stars mark the positions of the pulsars we use
in our fiducial configuration. The ones with black outline indicate the 42
pulsars not present in \emph{IPTA}\_20.
The main properties (total mass, luminosity distance from Earth,
frequency of their GW signal, and mass ratio) of the selected
binaries are shown by the round markers in Figure \ref{fig:binprops},
which are color-coded as in Figure \ref{fig:binposits}.
We discuss the distribution and the properties of the selected
binaries in further detail in Section \ref{sec:res}.

For the sub-set of binaries that can be localised in \emph{IPTA}\_25
(highlighted with the colored round markers in the figures above),
we also explore how their properties and localisation area are
constrained in \emph{IPTA}\_30.

\subsection{Estimation of binary parameters and host properties}
For each population in which the loudest source is consistent
with the criteria described above (i.e. $S/N\geq8$ in \emph{IPTA}\_25
and  $S/N<4$ in NG15) we use the \texttt{enterprise} software package
\citep{Ellis20} to inject a realization of common uncorrelated red
noise which represents the GW background, following the process
described in Section \ref{sec:gwselec}, and a continuous wave source
with the properties of the loudest binary of the respective population.
To estimate the detectability of the sources, the times of arrival
(TOAs) are simulated for all 3 PTA configurations we describe in
Section \ref{sec:ptaconfigurations}. For the subset of binaries with
$S/N\geq8$ in \emph{IPTA}\_25, we perform an all-sky search for a
continuous wave, using the \texttt{QuickCW} package \citep{Becsy22}.
The same analysis is then repeated for the same subset of binaries in
\emph{IPTA}\_30. The pipeline estimates posterior distributions for
the binary parameters, including its sky coordinates, thus allowing
us to constrain the size and shape of A90. In particular, we run
parallel tempered Monte Carlo Markov Chains (MCMCs) with 10 individual
chains, a maximum temperature parameter of 5, a total of 1 billion
steps, and store one step every 1,000. We use log-uniform priors for
the frequency $f$: $[0.5, 100]$ nHz, the chirp mass $\mathcal{M}$:
$[10^7, 10^{11}]M_\odot$, the strain $h_0$: $[10^{-18},10^{-11}]$,
and the amplitude of the GW background $A_{\rm GWB}$: $[10^{-20},10^{-11}]$.
We use uniform priors for the GW background spectral index
$\gamma_{\rm GWB}$: $[0, 7]$, the phase of the binary's orbit:
$[0, 2\pi]$, the polarization angle: $[0, \pi]$, and the cosine of the
inclination between the orbital angular momentum and the line of sight:
$[-1, 1]$. For the two angular coordinates on the sky, we use uniform
priors for the cosine of the polar angle: $\cos(\theta)\in[-1, 1]$ and
the azimuthal angle: $\phi\in[0, 2\pi]$. These angular coordinates are
then converted to equatorial coordinates, right ascension RA = $\phi$
and declination Dec = $\pi/2-\theta$. The posterior distribution of each
source parameter is then represented by the samples of the chain with
the lowest temperature, with the first 25 percent of the samples discarded
as \emph{burn-in}.

In order to produce a well-defined map of the sky localisation, we
associate each posterior sample of the sky coordinates to a pixel
of an Healpix map with NSide = 32. This mapping divides the sky in a
total of 12,288 pixels of equal area ($\approx3.36\ {\rm deg}^2$).
We sort the pixels in descending order based on the number of posterior
samples they contain. The localisation area, A90, is then defined
as the ensemble of pixels that contain 90 percent of the total number
of samples. We calculate the size of the localisation area, A90, by
multiplying the number of pixels by the pixel area. We then identify
potential hosts by cross-matching this localisation area with two
all-sky catalogues containing ETGs and AGN (see below for details).

To incorporate the EM information of the galaxy catalogues in our host
search, we convert the posterior distributions obtained with
\texttt{QuickCW} into distributions of EM observables. The chosen
EM quantities depend on the information that is available in each
galaxy catalogue. In particular, the ETG catalogue provides values
of the redshift, which can be converted into luminosity distance
$d_{\rm lum}$, and of the stellar mass of the galaxy, which we convert
into the mass of the central SMBH and associate to the binary total
mass, $M_{\rm tot}$. On the other hand, the AGN catalogue lists the
apparent magnitudes in the W1 band, $m_{\rm W1,AGN}$, for each source.

\texttt{QuickCW} provides posterior distributions for $\mathcal{M}$,
$f$, and $h_0$ of the source (in addition to sky coordinates, inclination,
GW phase, and polarization angle). We use the above 3 quantities to
create a 2-dimensional posterior distribution of the total mass and
luminosity distance $p\left(M_{\rm tot},d_{\rm lum}\right)$, which can
be linked to the quantities available in the EM catalogues. For this,
we employ a rejection sampling method, like in \citet{Petrov24} and
\citet{Veronesi25}. 

First, we calculate the distribution of luminosity distance, $d_{\rm lum}$
from the posterior samples of \texttt{QuickCW}:
\begin{equation}
\label{eq:dl}
d_{\rm lum}=\frac{2\left({\rm G}\mathcal{M}\right)^{5/3}
\left(\pi f\right)^{2/3}}{c^4h_0},
\end{equation}
where G is the gravitational constant and $c$ is the speed of light
in vacuum. Then we linearly interpolate the resulting 2D probability
density distribution $p\left(\mathcal{M},d_{\rm lum}\right)$, without
marginalizing over $f$ and $h_0$.

We then randomly sample a value for the parameters $\mathcal{M}$,
$f$, and $h_0$, drawing from the same priors as in the MCMC analysis
to calculate a random draw in luminosity distance, $d_{\rm lum,rand}$,
from Equation \ref{eq:dl}. We then calculate a value of the chirp mass,
$\mathcal{M}_{\rm rand}$, drawing from a uniform distribution of the
mass ratio $q$, $q$:[0.1, 1], and a uniform distribution for the total
mass $M_{\rm tot}$. For the latter, the boundaries are set by
combining the prior boundaries for $q$ and $\mathcal{M}$, using the
definition of the chirp mass:
\begin{equation}
\label{eq:chirp}
   M_{\rm tot}=\mathcal{M}\frac{q^{-3/5}}{\left(1+q\right)^{-6/5}}.
\end{equation}
This ensures that the random values $M_{\rm tot,rand}$ are extracted
from a distribution that covers the entire range derived from the
ranges of the two mass-related parameters used in the MCMC analysis.

We then draw a random number, $N_{\rm rand}$, between 0 and the maximum
value of the interpolated $p\left(\mathcal{M},d_{\rm lum}\right)$, and
reject the random sample
$\left(\mathcal{M}_{\rm rand},d_{\rm lum,rand}\right)$, if
$N_{\rm rand}>p\left(\mathcal{M}_{\rm rand},d_{\rm lum,rand}\right)$.
We repeat this until the number of non-rejected samples is equal to the
number of posterior samples from \texttt{QuickCW}. The result is a joint
$p\left(M_{\rm tot},d_{\rm lum}\right)$ posterior composed of all the
non-rejected $d_{\rm lum,rand}$ and $M_{\rm tot,rand}$ samples. As
mentioned above, we can directly link these to quantities in the ETG
catalogue, and thus we can use the above joint distribution to rank
the ETGs within the localisation area. 

However, in order to perform a similar ranking for the AGN catalogue
we present in Section \ref{sec:agncat}, we convert
$p\left(M_{\rm tot},d_{\rm lum}\right)$ into a distribution of apparent
magnitude in the W1 band, $p\left(m_{\rm W1,AGN}\right)$, taking into
account the uncertainties and the scatter of the relations we use in
the conversion. We do so by first estimating the Eddington luminosity
$L_{\rm Edd}$ of an AGN with a mass of $M_{\rm tot}$:
\begin{equation}
    L_{\rm Edd}\sim 1.26\cdot10^{38}\frac{{\rm M}_{\rm tot}}
    {{\rm M}_{\odot}}{\rm erg\ s}^{-1}.
\end{equation}
For each sample of total mass, we also draw a value for the Eddington
fraction $\lambda_{\rm Edd}$ from distributions derived from \citet{Wu22}.
This catalogue uses quasar spectra from the Sloan Digital Sky Survey Data
Release 16 \citep{Lyke20} and provides estimates for properties, such as
$\lambda_{\rm Edd}$, and the SMBH mass, which here we associate to the
total mass of the binary, $M_{\rm tot}$. For each value of $M_{\rm tot}$,
we select the quasars from \citet{Wu22} with an estimated SMBH total mass
$\log_{10}M_{\rm tot}\pm0.05$ and randomly draw the value for
$\lambda_{\rm Edd}$ of one of these objects to calculate the bolometric
luminosity:
\begin{equation}
    L_{\rm bol}=\lambda_{\rm Edd}\cdot L_{\rm Edd}.
\end{equation}

Following \citet{Veronesi25}, we combine the spectral energy distributions
from \citet{Shang11} with information regarding quasar bolometric
luminosities and redshifts from \citet{Runnoe12} to obtain a distribution
of bolometric corrections in the W1 band. We then draw one sample from this
distribution and multiply the bolometric luminosity $L_{\rm bol}$ by it to
obtain the W1 luminosity $\nu_{\rm W1} L_{\nu_{\rm W1}}$ of the potential AGN
host. From this luminosity, we then calculate the absolute magnitude
$M_{\rm W1}$ using the same standard process as in Section 3.2 of
\citet{Veronesi25}. Finally, we calculate the apparent magnitude using the
value of $d_{\rm lum}$ which corresponds to the same posterior sample of the
$M_{\rm tot}$ value used to calculate $M_{\rm W1}$:
\begin{equation}
    m_{\rm W1}=M_{\rm W1}+5\log{\left(\frac{d_{\rm lum}}{\rm Mpc}\right)
    +25+0.171\cdot A_{V}},
\label{eq:abstoappmag}
\end{equation}
where $A_V$ is the dust extinction in the visible band, which we multiply
by 0.171 to convert it to that in the W1 band \citep{Cardelli89}.
Analogously to \citet{Veronesi25}, we use $A_V\approx0.19$ as the fiducial
value, corresponding to the median value of the extinction across the
sky obtained from the map presented in \citet{Chiang23}.

\subsection{Galaxy catalogues}
We cross match the GW localisation areas with two all-sky galaxy catalogues,
one containing only ETGs and the other containing only AGN. These catalogues
are used in parallel in our analysis. Below we briefly describe their
construction and essential EM properties, that we take into account in
the host ranking system.

We emphasize that the simulated binaries have a luminosity distance
from Earth dictated by the population synthesis models described
in Section \ref{sec:popcreat} and they are placed in a random sky
location during the binary selection process in Section \ref{sec:gwselec}.
Therefore, the binary is not associated with any real galaxy in any of
the two EM catalogues. However, our goal here is to assess the number of
potential hosts in future multi-messenger searches in a realistic
scenario, i.e., examining a binary from a realistic binary population,
detected and localised with one of the standard PTA pipelines and
cross-matched with all-sky catalogues of high completeness, while
incorporating uncertainties both from the GW analysis and the EM
catalogues. This also allows us to estimate what fraction of the GW
sky maps lies within the footprint of such EM surveys, and assess
our ability to rank the candidate hosts based on their EM properties.
In future studies, we will repeat this analysis injecting the selected
binaries in true galaxies, and we will check how high the true host is
ranked with respect to the sorted list of all potential hosts, therefore
estimating the accuracy of the ranking method. In this work we estimate
its precision.

\subsubsection{The ETG catalogue}
\label{sec:etgcat}
The ETG catalogue is composed of the Two Micron All Sky Survey Photometric
Redshift catalogue (2MPZ, \citealt{Bilicki14}), and of the WISExSuperCOSMOS
Photometric Redshift catalogue (WISExSCOS, \citealt{Bilicki16}). It contains
19,586,966 photometrically selected galaxies, each with an estimated value
of photometric redshift and stellar mass. Every object was detected both by
the Wide-field Infrared Survey Explorer (WISE, \citealt{Wright10}) and by
the SuperCOSMOS Sky Survey \citep{Hambly01}, but the brightest objects (i.e.
the ones with an infrared magnitude $m_{\rm W1}\leq13.8$), which were also
detected by the Two Micron All Sky Survey (2MASS, \citealt{Cutri03}), are
contained in 2MPZ and not in WISExSCOS.

The combination process and the estimation of the stellar mass of each
galaxy is described in detail \citet{Pilawa26}. In brief, the stellar
mass is calculated from its intrinsic luminosity in the W1 band, using
the correlation between the mass-to-light ratio and the W1-W2 color from
\citet{Cluver14}. We use the stellar mass estimates to calculate the mass
of the SMBHs (which we associate with the total mass of the binary),
assuming that all the stars are contained in the bulge. For this, we use
the scaling relation of Equation \ref{eq:mmbulge}, adopting in this case
the best-fit parameters calculated in \citet{McConnell13}: $\mu=8.46\pm0.08$,
$\alpha_\mu=1.05\pm0.11$, and intrinsic scatter $\epsilon_\mu=0.34$.
We note that the estimates on the galaxy stellar mass obtained from the
W1 magnitude are found to be systematically lower ($\approx0.3$ dex) than
the ones obtained from the magnitude in the K band for the objects that
have a detection both in WISE and in 2MASS \citep{Pilawa26}. Future versions
of the ETG catalogue we use, not available at the time of writing, will
contain stellar mass estimates for all the galaxies corrected taking into
account this discrepancy, since the K magnitude is considered to be the
one that best relates to such galaxy property. Higher estimates of stellar
masses will lead to higher estimates of total binary masses. This will not
change the number of potential host candidates, but will moderately decrease
the rejection rate of ranking methods such as the one presented in this work,
since fewer galaxies will be found incompatible with the typically large
values of $M_{\rm tot}$ estimated from the GW pipelines.

The ranking system presented in Section \ref{sec:ranking} requires
an estimate of the uncertainty for each employed quantity, i.e. the
redshift and the mass of the central SMBHs for this catalogue. To
estimate the uncertainty $\sigma_z$ on the photometric redshift, we
use the prescriptions from \citet{Turski23}. Specifically, we use
\(\sigma_z=0.052\cdot z+0.008\), if the parent catalogue of the object
is 2MPZ, and \(\sigma_z=0.085\cdot z+0.019\), if it is in WISExSCOS.
For the uncertainty on the SMBH mass, $\sigma_{M_{\rm tot}}$, we use
standard rules of error propagation:
\begin{equation}
\begin{aligned}
\sigma_{M_{\rm tot}}^2 &= \mu^2\sigma_{M_{\rm b}}^2
+\sigma_{\mu}^2\left(M_{\rm b}-11\right)^2 \\
&\quad + \sigma_{\alpha_\mu}^2 + \epsilon_\mu^2,
\end{aligned}
\end{equation}
where $\sigma_{\mu}$ and $\sigma_{\alpha_\mu}$ are the uncertainties
of the best fit parameters in the scaling relation of \citet{McConnell13},
and $\sigma_{M_{\rm b}}$ is the intrinsic scatter of the correlation used
to calculate the bulge mass, which here we set to $\sigma_{M_{\rm b}}=0.1$
based on Fig. 6b of \citet{Cluver14} \footnote{The value of
$\sigma_{M_{\rm b}}$ is not explicitly stated in \citet{Cluver14}, and
therefore we obtain an estimate of 0.1 based on visual inspection of the
aforementioned Figure 6b. Small deviations from this value are not expected
to significantly affect our results.}.

\begin{figure}
    \centering
    \includegraphics[trim= 50 0 50 0
    ,clip,width=0.98\columnwidth]{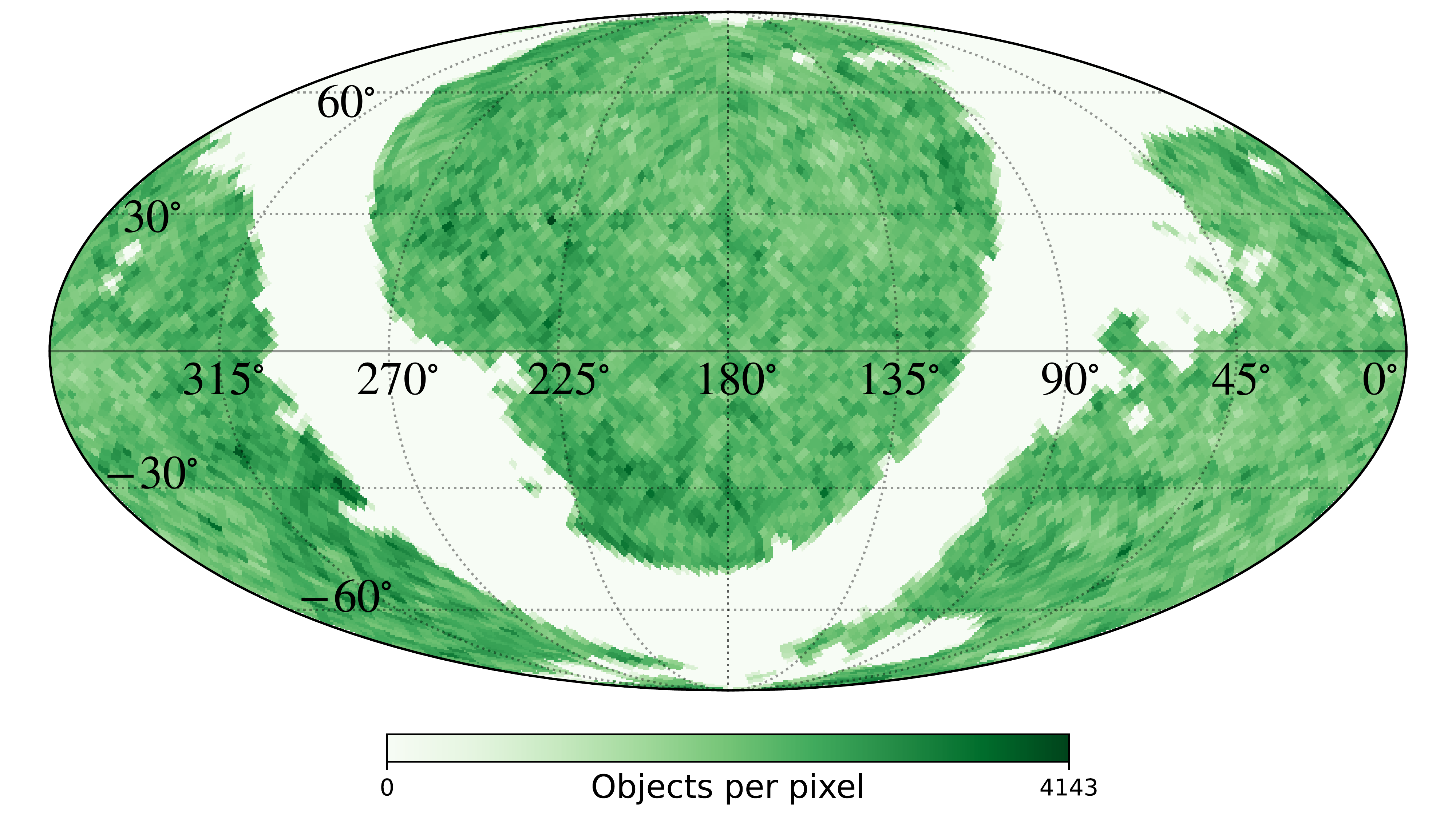}
    \includegraphics[trim= 50 0 50 0
    ,clip,width=0.98\columnwidth]{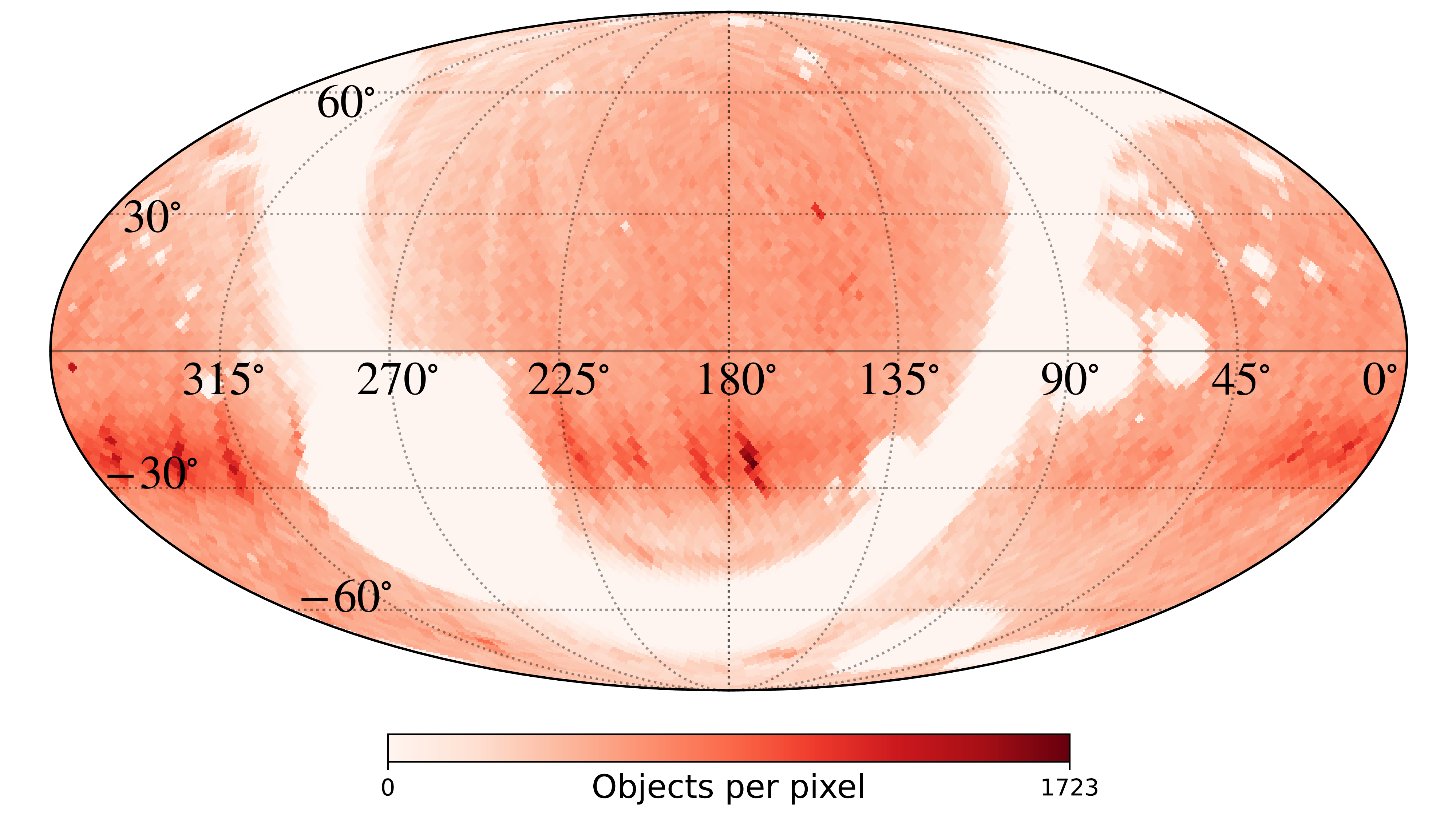}
    \caption{Mollweide projection of the all-sky catalogues of
    ETGs (upper panel) and AGN (lower panel ) used in this work.
    The colour of each pixel denotes the number of objects it
    contains in an Healpix projection with NSide=32.}
    \label{fig:cats}
\end{figure}

We note that 2MPZ covers some areas that are avoided in WISExSCOS,
especially in regions with low Galactic latitude. To guarantee
uniformity when cross-matching with GW sky maps, we exclude the
regions covered by 2MPZ but not by WISExSCOS, which are highly
incomplete. In practice, this corresponds to rejecting sky regions
where the pixels of the Healpix map with NSide = 32 have fewer than
415 objects. After excluding these areas, we are left with a total
of 19,364,973 ETGs, the sky distribution of which is shown in the
upper panel of Figure \ref{fig:cats}.

\subsubsection{The AGN catalogue}
\label{sec:agncat}
From the ETG galaxy catalogue candidate AGN have been discarded
(through color-based selection), so we use a complementary catalogue
to include potential AGN hosts. In this work, we remain agnostic
about the \textit{a priori} probability of a PTA source being
associated with a specific type of host \citep[e.g., see][]{Truant26},
and perform our analysis for the ETG and AGN catalogues in parallel.

We use an AGN catalogue from \citet{Assef18}, which contains a total of
4,543,530 objects detected by WISE, present in the AllWISE Data Release.
For each entry the catalogue contains the measured apparent magnitudes
the infrared bands of WISE, and their photometric uncertainties. They
are identified as AGN candidates with 90 percent reliability (comprising
the so-called R90 catalogue) using a cut based on the $W1$ and $W2$
magnitudes:
\begin{equation}
    {\rm W1-W2}>0.662\cdot e^{0.232\left({\rm W2}-13.97\right)^2}.
\end{equation}
The R90 catalogue has a high level of uniformity and reliability.
Typically, such catalogues have to compromise on completeness to achieve
high reliability, but since we are interested in massive binaries,
which live in luminous hosts, completeness is likely not a significant
limitation. In the lower panel of Figure \ref{fig:cats}, we show the
sky distribution of the AGN in this catalogue.

The over-dense regions in the southern hemisphere are caused by the
survey strategy which was adopted to correct for the decreased
sensitivity in the south Atlantic anomaly (see Section 3.3.2 of
\citealt{Assef18} for details). These over-densities are present
only for faint sources (with a $S/N<10$ in the W2 band) and their
inclusion has therefore minimal impact on our results, but a future
study may consider a more careful selection.

We note that when comparing the total number of objects in the AGN
and ETG catalogues, we find the former to be $\approx23.5$
percent of the latter. This number is higher than the typical AGN
fraction \citep{Kauffmann03}, because R90 is deeper than the ETG
catalogue, with the former containing objects with $m_{\rm W1}>19$,
and the latter imposing a limiting W1 magnitude of 17 to achieve 
higher uniformity \citep{Bilicki16}.

\subsection{Host ranking system}
\label{sec:ranking}
In this section, we present the system we develop to rank the potential
host galaxies. To each galaxy (ETG or AGN, separately) within the
localisation area, we associate a host score $\mathcal{H}$, which we
define as follows:
\begin{equation}
    \mathcal{H}=p_{\rm sky}\cdot p_{\rm cat},
\end{equation}
where $p_{\rm sky}$ and $p_{\rm cat}$ are probability density values
that estimate how well the candidate host properties match the posteriors
of the detected PTA source. The factor $p_{\rm sky}$ measures how well
the host sky position matches the GW sky map, and $p_{\rm cat}=\prod_xp_x$
is its equivalent for the other host properties $x$ we consider. In
particular, for ETGs the additional information we use is the estimated
total mass of their SMBH and their luminosity distance calculated from
the redshift, ($p_{\rm cat,ETG}=p_{M_{\rm tot}}\cdot p_{d_{\rm lum}}$),
while for AGN we rely on their apparent W1 magnitude
($p_{\rm cat,AGN}=p_{m_{\rm W1,AGN}}$), since in state-of-the-art AGN
all-sky catalogues with homogeneous completeness (necessary for this type
of analyses), the total mass of the central SMBH and the redshift are not
separately listed. The value of $p_x$ is proportional to the probability
of measuring the two distributions of the property $x$ (one from the
GW analysis and one from the EM catalogue) assuming that the underlying
true value of $x$ is the same for both of them.

We calculate $p_{\rm sky}$ for each candidate host by first identifying
the pixel within A90 in which the galaxy is located (keeping the resolution
of NSide=32 throughout the analysis). Then we calculate the probability
density of the PTA source being associated with that pixel, $p_{\rm sky}$,
dividing the number of posterior samples of that pixel by the total number
of samples.

While the derivation of $p_{\rm sky}$ is straightforward, since
the uncertainty in the sky coordinates of the galaxies is negligible,
the uncertainty on the redshift and on the EM observables needs to
be accounted for in calculating $p_x$, and thus requires a different
approach. For each property $x$, we calculate $p_x$ for each potential
host by first creating a difference distribution composed of 10,000
samples of the quantity $\Delta_x$. Each sample is the difference
between one random value drawn from the GW-posterior-derived
distribution of the observable $x$ and one drawn from a Gaussian with
mean and standard deviation taken from the EM catalogue for the specific
candidate host under consideration.

We approximate the above probability density function,
$p\left(\Delta_x\right)$, with a kernel density estimation and then
draw 1,000 samples to calculate the probability to exceed,
${\rm PTE}_x$, as:
\begin{equation}
    {\rm PTE}_x=\sum_{i=1}^{1000}\Theta\left(p\left(\Delta_x=0\right)-
    p\left(\Delta_{x,i}\right)\right),
\end{equation}
where $\Theta$ indicates the Heaviside step function (which returns 0 for
negative values, and 1 for positive ones), and $\Delta_{x,i}$ is the $i$-th
random draw from $p\left(\Delta_x\right)$. The value of ${\rm PTE}_x$
therefore denotes the probability of extracting from $p\left(\Delta_x\right)$
a value of $\Delta_x$ that has an associated probability density higher
than the one associated to $\Delta_x=0$. If the underlying value of $x$
was the same for the two distributions, $p\left(\Delta_x\right)$ would
peak at zero, and for this reason ${\rm PTE}_x$ would be null. However,
for different starting distributions, ${\rm PTE}_x$ will be positive. The
higher the value of ${\rm PTE}_x$, the more the two starting distributions
differ. 

We then use the value of ${\rm PTE}_x$ to define a sigma equivalent
$\sigma_{{\rm eq},x}$. An intuitive definition of this quantity is the
following: if the galaxy under consideration is the true host,
the observed difference in the two distributions of the observable $x$
(i.e. the one derived from the GW posteriors and the Gaussian parametrized
by the mean and the standard deviation from the catalogue or calculated
as in Section \ref{sec:etgcat}) is as likely to obtain as a random draw
from a Gaussian distribution with distance from the mean equal to a
number $\sigma_{{\rm eq},x}$ of standard deviations. This quantity is
calculated by inverting the following relation:
\begin{equation}
    {\rm PTE}_x=p\left(|Z|\geq\sigma_{{\rm eq},x}\right),
\end{equation}
where $Z$ is a random draw from a Gaussian distribution of null mean
and a standard deviation of 1. The value of $\sigma_{{\rm eq},x}$ is
therefore the one that has a probability of ${\rm PTE}_x$ to be smaller
(in absolute value) than a random draw of the function $Z$ is extracted
from. Inverting the equation therefore leads to:
\begin{equation}
    \sigma_{{\rm eq},x}=\sqrt{2}\cdot{\rm erf}^{-1}(1-{\rm PTE}_x),
\end{equation}
where erf denotes the error function. 

We can use $\sigma_{{\rm eq},x}$ to calculate the probability density
$p_{x}$, as the probability density associated to it assuming a normal
Gaussian distribution:
\begin{equation}
    p_{x}=\sqrt{\frac{2}{\pi}}e^{-\frac{\sigma_{{\rm eq},x}^2}{2}}.
\end{equation}
Putting everything together, for every ETG within the localisation area
A90, we calculate the ETG host score as:
\begin{equation}
    \mathcal{H}_{\rm ETG}=p_{\rm sky}
    \cdot e^{-\frac{\sigma_{{\rm eq},M_{\rm tot}}^2}{2}}
    \cdot e^{-\frac{\sigma_{{\rm eq},d_{\rm lum}}^2}{2}},
    \label{eq:scoreetg}
\end{equation}
and similarly for every AGN within the same sky region, we calculate
the AGN host score as:
\begin{equation}
    \mathcal{H}_{\rm AGN}=p_{\rm sky}
    \cdot e^{-\frac{\sigma_{{\rm eq},m_{\rm W1,AGN}}^2}{2}},
    \label{eq:scoreagn}
\end{equation}
We drop the $\sqrt{\frac{2}{\pi}}$ factor, because it is common
to all ETGs and AGN. We emphasize that we are constructing a ranking
system that assigns a score to every potential host within A90 to rank
them. However, the values of these scores do not correspond to an
absolute probability density of an object being the true host. Moreover,
the ranking is relative within each catalogue, because the two scores
have been calculated separately and have different normalizations.
This means that if an ETG has a higher score with respect to an AGN,
it does not necessarily imply that the former is more likely to host
the binary with respect to the latter. The only factor that allows
a direct comparison between the two host types is $p_{\rm sky}$.

Finally, to quantify the efficiency of the ranking system, for each
localised binary, we calculate the ratio between the 95-th percentile
of the host score distribution and its median. We further refer to this
efficiency parameter as $\mathcal{E}_{95-50}$, and its value expresses
how much more likely the top performing candidates are to be the true
host, compared to the typical candidate within A90. A higher value of
$\mathcal{E}_{95-50}$ therefore indicates a better ability to select
the best candidates, and a higher efficiency of the ranking system.
This value is calculated for ETGs and AGN, separately, for the
well-localised binaries both in \emph{IPTA}\_25 and \emph{IPTA}\_30.

%------------------------------------------------------------------
%------------------------------------------------------------------
%------------------------------------------------------------------

\section{Results}
\label{sec:res}

\subsection{Selection of detectable and localisable binaries}

We create 1,000 realizations of binary populations and select the
loudest binary for each population. We place these binaries in random
sky positions and examine their detectability in \emph{IPTA}\_20
and \emph{IPTA}\_25. We select binaries that have $S/N\geq4$, while
not being detectable in NG15. We find that a total of 212 and 378
binaries pass this detectability selection for \emph{IPTA}\_20 and
\emph{IPTA}\_25, respectively. Therefore, given our simulations of
upcoming PTA sensitivities and our models for the binary population,
we estimate the probability of resolving a binary in an upcoming IPTA
dataset with a baseline of 20 and 25 years to be 21.2 and 37.8
percent, respectively. However, we emphasize that the S/N threshold
is only a proxy for the detection statistics employed in PTA searches,
and these probabilities should be interpreted only as rough estimates.

The binaries that pass the detection threshold in \emph{IPTA}\_25 are
highlighted with blue squares in Figures \ref{fig:binposits} and
\ref{fig:binprops}, while the remaining population of loudest binaries
is shown with gray squares. We see that the detectable sources are
isotropically distributed on the sky and have properties (distance
and total mass) statistically indistinguishable from the overall
population. Specifically, the 25$-th$, 50$-th$, and 75$-th$ percentiles
for the luminosity distance are
$\log_{10}\left(d_{\rm lum}/{\rm Mpc}\right)=2.65_{2.28}^{2.96}$, where
the main reported value indicates the median, while the subscript and
the superscript the two other quartiles. The same percentiles for the
total binary mass are
$\log_{10}\left(M_{\rm tot}/M_\odot\right)=9.72_{9.50}^{9.95}$.
These sources tend to have low frequencies, with $\approx64$ percent of
the detectable binaries distributed in the first 5 of the 14 frequency
bins. A similar result is found for \emph{IPTA}\_20, where $\approx69$
percent of detectable binaries are in that frequency range.

The above values for the detectable binaries are broadly in agreement
with findings in previous studies \citep{Rosado15,Becsy22b,Gardiner25}.
This is true even if the population synthesis models and the employed
selection criteria are different among studies. We acknowledge that we
find a non-negligible fraction of detectable binaries with masses comparable
with (or higher than) $10^{10}M_\odot$. SMBHs with such high masses are
present in the local Universe, albeit relatively rare. For example, the
MASSIVE survey \citep{Ma14} has identified 116 ETGs with stellar masses
bigger than $4\cdot10^{11}M_\odot$ within 108 Mpc from Earth \citep{Veale18}.
For this reason, we choose not to discard these most massive systems from
our analysis, since they were produced by self-consistent population
synthesis models, in agreement with the measured GW stochastic background. 

Next, we explore how many of these binaries reach a $S/N\geq8$, which is
the threshold we set for potentially being able to constrain their sky
position and trigger a multi-messenger search (see Section \ref{sec:gwselec}).
We find that of the 1,000 binary populations we generated, 3 and 38 would
produce a binary that passes this cut in \emph{IPTA}\_20 and \emph{IPTA}\_25,
respectively. The small number of binaries localisable in \emph{IPTA}\_20
is not surprising given the S/N thresholds we set; the difference in
sensitivity between NG15 and \emph{IPTA}\_20 is limited and thus detecting
a source with a well-constrained A90 in the third data release of IPTA
is unlikely. Moreover, since the probability of resolving a binary in
\emph{IPTA}\_20 is very small (less than one percent), we focus on
\emph{IPTA}\_25 for the remaining analyses and consider it our fiducial
PTA.

In Figure \ref{fig:binposits}, we show the sky positions of the
potentially localisable binaries. As mentioned in Section
\ref{sec:gwselec}, the size of each marker is proportional to the
total mass of the system, while its color denotes the S/N in the
\emph{IPTA}\_25 dataset. Unlike the detectable binaries above (with
S/N$\geq4$), the distribution of these sources is highly anisotropic
with about half of the binaries (20 out of 38) falling in a
well-constrained region of the sky, with a right ascension between
RA=$90^\circ$ and RA=$225^\circ$, and a declination between
Dec=$-45^\circ$ and Dec=$0^\circ$. This region has an area of
approximately 5470 ${\rm deg}^2$, corresponding to only $\approx13$
percent of the sky. This specific region is where the boost in
sensitivity between NG15 and \emph{IPTA}\_25 is highest, most likely
due to the addition of several pulsars in the southern hemisphere,
e.g., from MPTA and PPTA. We further discuss this in Section
\ref{sec:discus:prospects}.

The main properties of the 38  binaries that could be spatially
resolved in \emph{IPTA}\_25 are shown by the colored markers in
Figure \ref{fig:binprops}, which are color-coded according to the
S/N of their detection in our fiducial PTA configuration. The
25$-th$, 50$-th$, and 75$-th$ percentiles of the distribution of
their luminosity distance are
$\log_{10}\left(d_{\rm lum}/{\rm Mpc}\right)=2.42_{2.07}^{2.67}$.
The same percentiles for the distribution of the total mass of
these sources are 
$\log_{10}\left(M_{\rm tot}/M_\odot\right)=9.79_{9.61}^{10.01}$.
This means that the selected sources are marginally closer to
Earth and more massive than the rest of the population. We conclude
that the first localisable binary will likely be relatively nearby
(most binaries are roughly within 1Gpc, with the sole exclusion of
2 sources) and very massive (half of the sources have a mass equal
or greater than $10^{9.8}M_{\odot}$). In addition, all of the
potentially localisable sources are in the first six frequency bins,
as is evident from the upper right panel of Figure \ref{fig:binprops}.
The resolvable binaries tend to have relatively high mass ratios,
typically $q>0.4$, as shown in the lower panel.

\subsection{Binary parameters, localisation areas, and galaxy cross-matches}

For the 38 binaries which are potentially localisable with
\emph{IPTA}\_25, we follow a more careful analysis. We simulate
TOAs with the GW background from the respective population and
inject one continuous-wave source, with properties as described
above. Then we run the detection pipeline \texttt{QuickCW} to
constrain the posterior distributions of their parameters. This
allows us to constrain the localisation area, which we then use
to cross-match with the two galaxy catalogues. Of those, 30 have
a size of A90 smaller than 2,000 ${\rm deg}^2$ ($\approx$ 5 percent
of the total sky area), while the remaining 8 are poorly localised
and therefore excluded from the rest of our analysis. This
highlights that, even though we set a threshold of $S/N=8$ to
select sources with well-constrained sky positions, not all
injections return a 90 percent localisation area small enough
to allow for an efficient host search. This is in line with
findings from \citet{Petrov24}, where only 5 of the 9 binaries
with $S/N=8$ have a constrained localisation area. Factors beyond
the S/N, like the chirp mass of the binary and its position
relative to the pulsars of the array (e.g., proximity to the
best-timed pulsars or to those with well-known distances from
Earth) play an important role in constraining the binary sky
location \citep{Petrov24,Taylor26}. The three quartiles of the
distribution of the size of A90 for the well-localised binaries
are ${\rm A90}=408_{250}^{509}{\rm deg}^2$. 

We then calculate the overlap between the GW sky maps and the EM
catalogues and quantify the coverage fraction $f_{\rm cover}$. We
find that A90 is entirely within the footprint of both EM catalogues
($f_{\rm cover}=1$) for 8 of the 30 well-localised sources, while in
one case we get $f_{\rm cover}=1$ for the AGN catalogue but not for
ETGs. Therefore, in approximately two thirds of the cases, the
GW localisation area is not fully covered by the EM surveys. This can
significantly limit our ability to find the true host of PTA sources.
On the other hand, there are no cases in which A90 is completely
contained in a region of the sky with no support from the catalogues
($f_{\rm cover}=0)$, likely because the typical size of A90 constrained
by our fiducial PTA configuration is more extended than the gaps of the
catalogues, e.g., the Galactic plane (see Section \ref{sec:constraintsimprov}
for a discussion on the difference in A90 constrained by \emph{IPTA}\_25
and \emph{IPTA}\_30). The number of potential hosts reported below are
not normalized based on $f_{\rm cover}$, but we separately use the
coverage factor to estimate the number of missing hosts in Section
\ref{sec:missing_gals}. This work focuses on providing realistic
expectations regarding how many observed galaxies will have the
possibility of being promptly followed-up in realistic host searches,
and on ranking them. If the GW localisation area extends to sky regions
not covered by existing surveys, this will motivate new observational
campaigns, and the newly detected galaxies will be subsequently added
to the list of potential hosts and ranked.

Cross-matching the GW maps with the ETG and AGN catalogues we calculate
the number of potential hosts. The three quartiles of the distribution
of the number of ETGs within A90 for the 30 well-localised binaries
are $N_{\rm ETG}=189,286_{95,290}^{312,231}$, while for the AGN hosts
they are $N_{\rm AGN}=39,655_{11,786}^{76,931}$. The fact that these
numbers are much larger than the ones obtained in \citet{Petrov24},
where the number of potential hosts within A90 are between 285 and
1,238 for the 5 localised binary detections with $S/N=8$, is
explained by the much deeper catalogues used in this work, since the
sizes of the localisation areas are not significantly different. The
catalogue used in \citet{Petrov24} contains a total of 43,532 entries.

The number of potential hosts is directly linked to size of the
localisation area. However, the other main binary parameters ($f$,
$h_0$, and $\mathcal{M}$) also determine our expectations about the
binary hosts, since they are used to calculate the distances of the
binaries using Equation \ref{eq:dl}, and the expected distributions
of their EM properties, which in turn affect the ranking, the results
of which are presented in Section \ref{sec:resranking}. For this
reason, we also quantify how well these properties are constrained by
\texttt{QuickCW}. We find that the frequency and the strain of the
30 well-localised binaries are correctly constrained around the true
value. In particular, we find that the percent error of the median of
the posterior compared to the true injected value of the frequency and
strain in log space is on average 0.1 and 0.7 percent, respectively.
The interquartile range (between the first and the third quartile) of
the posterior distribution in log space is on average 0.02 for the
frequency and 0.20 for the strain.

On the other hand, we find that the chirp mass is in general poorly
constrained. The interquartile range of its posterior distribution
in log space is on average 1.27. The fact that the PTA data
cannot provide tight constraints on the chirp mass has important
implications in the search for the host galaxy. This well-known
limitation \citep{Sesana10,Charisi24} is mainly caused by the fact
that PTAs detect binaries early in their inspiral stage, where the
frequency evolution is slow, and not during the chirping that happens
right before the merger, which is the most critical stage for
constraining the mass of the binary and its luminosity distance.
This poorly constrained mass parameter is in turn, translated into
poorly constrained distributions for $d_{\rm lum}$, $M_{\rm tot}$,
and $m_{\rm W1,AGN}$. In \citet{Veronesi25}, we demonstrated that
the uncertainty in the chirp mass contributes to $\approx98$ percent
of the uncertainty on the photometric properties of the host. Only
the remaining $\approx2$ percent comes from uncertainties in the EM
observables (e.g., the intrinsic scatter in the scaling relationship
between the binary mass and the bulge mass of an ETG host, or the
distribution of Eddington ratios for a given source mass in the
case of an AGN).

\subsection{Constraints improvement with \emph{IPTA}\_30}
\label{sec:constraintsimprov}

Next we explore how our results evolve with the addition of 42
monitored pulsars and 5 years of baseline with respect to our
fiducial PTA configuration. With \emph{IPTA}\_30 we find a total
of 507 binaries to be detectable and 141 to be localisable,
while not being detectable by NG15. This finding suggests that in
the next ten years we have a $\approx50$ percent probability of
resolving the first binary, and a $\approx15$ percent chance of
being able to potentially localise it within a sky region small
enough to allow for host searches. These numbers likely represent
lower limits for our expectations, since upcoming observatories
like the Square Kilometre Array (SKA, \citealt{Dewdney09}) or the
Deep Synoptic Array (DSA, \citealt{Hallinan19}) will increase the
number of monitored pulsars faster than what we assume in this
work. The quartiles of the total mass distribution in logarithmic
space for the 141 potentially localisable binaries are
$\log_{10}\left(M_{\rm tot}/M_\odot\right)=9.76_{9.46}^{9.95}$.
The same values for the luminosity distance are
$\log_{10}\left(d_{\rm lum}/{\rm Mpc}\right)=2.57_{2.18}^{2.92}$,
and 101 of them are in the first 5 frequency bins. These sources
follow more closely the distributions of the full population
compared to the potentially localisable binaries in \emph{IPTA}\_25,
which are marginally more massive and nearby. Even though we still
find some evidence for spatial clustering of the localisable binaries
in the same region as in our fiducial configuration, the observed
concentration in the southern hemisphere is weaker and the 141
selected sources are generally distributed all over the sky. This
is to be expected, since with the increase in sensitivity with
respect to \emph{IPTA}\_25, the requirement of non-detectability
in NG15 is less stringent. 

However, the main goal of the \emph{IPTA}\_30 simulations is to
explore how the posteriors evolve with the addition of new pulsars
and the extension of the baselines for the 30 binaries that are
already well-localised with our fiducial configuration. Therefore,
we do not simulate the detection of all 141 potentially localisable
binaries.

Before presenting the results for the entire population, we
show the evolution of the size of A90 and the number of potential
hosts for one source. The binary we choose has a total mass of
$M_{\rm tot}=10^{9.05}M_\odot$, a mass ratio of $q=0.61$, a
luminosity distance from Earth of $d_{\rm lum}=37{\rm Mpc}$, a
frequency of $f=12{\rm nHz}$, and a S/N of 8.46 and 12.25 in
\emph{IPTA}\_25 and \emph{IPTA}\_30, respectively. This binary
was chosen for its large value of $f_{\rm cover}$, which is
greater than 0.9 for both EM catalogues. The size of A90 is above
average, resulting in a number of potential hosts on the higher
end of the distribution, allowing us to demonstrate our method
in a conservative scenario.

\begin{figure}
    \centering
    \includegraphics[trim= 400 100 400 100
    ,clip,width=0.98\columnwidth]{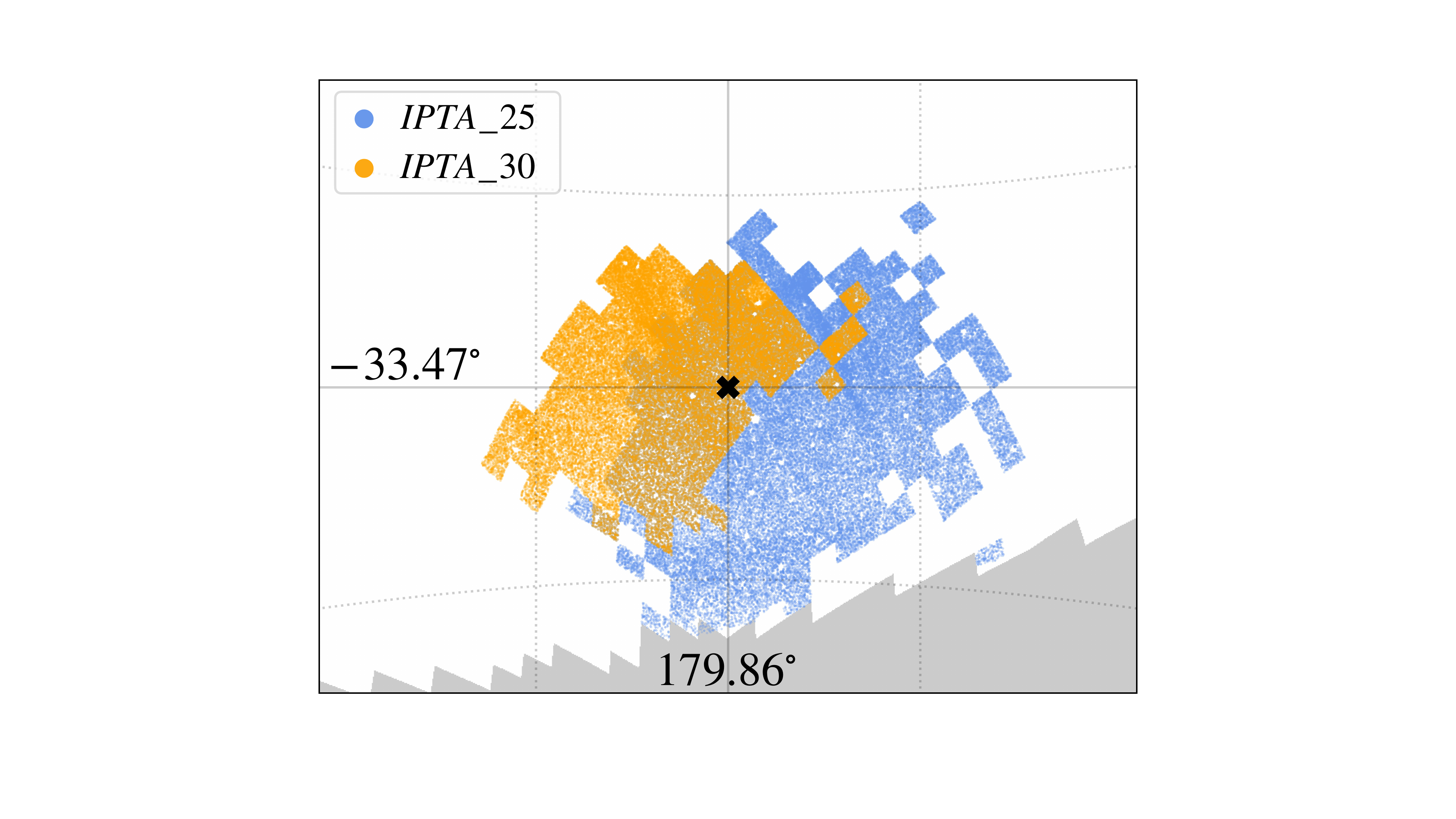}
    \caption{localisation area at the 90 percent credibility
    level for the example binary (the true position is marked
    by the black cross), as constrained with \emph{IPTA}\_25
    (blue) and \emph{IPTA}\_30 (orange). The localisation areas
    are delineated by the positions of the candidate AGN hosts
    from the R90 catalogue, since this binary has a high coverage
    factor ($f_{\rm cover}\approx1$) for both PTA configurations.
    The gray area denotes the region of the sky with a Galactic
    latitude of $|b|\leq10^\circ$.}
    \label{fig:binposits380}
\end{figure}

In Figure \ref{fig:binposits380}, we show the sky position and
localisation areas for the selected binary. In particular, we
mark with a black cross the position of this binary, while the
blue and orange dots are the positions of all the AGN within A90
as constrained by \emph{IPTA}\_25 and \emph{IPTA}\_30, respectively,
which practically delineate the respective A90s.
As in Figure \ref{fig:binposits}, the gray area marks the region of
the sky with a Galactic latitude of $|b|\leq10^\circ$. The size of
its localisation area decreases from $\approx645$ ${\rm deg}^2$ in
\emph{IPTA}\_25 to $\approx349$ ${\rm deg}^2$ in \emph{IPTA}\_30,
while the coverage factor with respect to the ETG and AGN catalogues
increases from $\approx91$ percent and $\approx96$ percent,
respectively, to full coverage $f_{\rm cover}=1$ in both catalogues.
The number of potential ETG and AGN hosts decreases from a total of
424,830 and 97,562 to 263,477 and 62,203, respectively. The number
of potential hosts decreases by $\approx40$ percent for both ETGs
and AGN. This percentage is slightly smaller than the 46 percent
decrease in localisation area. This is caused by the fact that
$f_{\rm cover}$ is higher in \emph{IPTA}\_30. We discuss the issue
of missing potential hosts in more detail in Section
\ref{sec:missing_gals}.

\begin{figure*}
    \centering
    \includegraphics[trim= 0 160 0 100
    ,clip,width=0.98\textwidth]{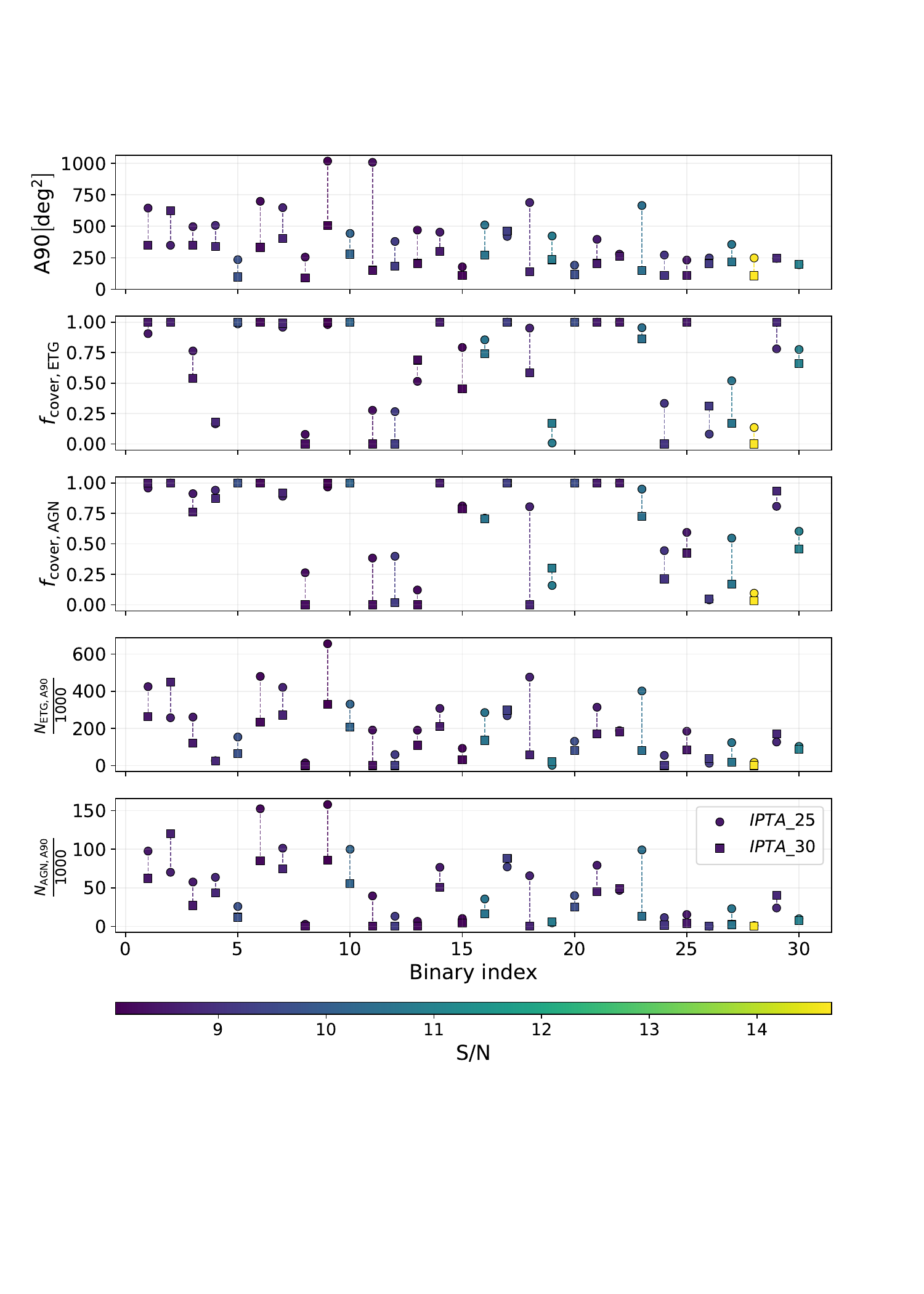}
    \caption{Evolution of the main results between \emph{IPTA}\_25
    (circles) and \emph{IPTA}\_30 (squares). The first panel shows
    the size of A90. The second and the third panels show the
    fraction of A90 that is in the footprint of the ETGs and the
    AGN catalogues, respectively. The last two panels show the
    number of potential hosts (ETGs and AGN) contained in A90.
    Each pair represents a separate well-localised simulated
    binary. In all the panels the simulated binaries are sorted
    on the horizontal axis in order of increasing total mass,
    color-coded by their S/N in \emph{IPTA}\_25 .}
    \label{fig:resultscrossmatches}
\end{figure*}

Figure \ref{fig:binposits380} demonstrates the reduction in
the size of A90, but also highlights that in general the
localisation area obtained with \emph{IPTA}\_30 is not
necessarily contained within the one obtained with
\emph{IPTA}\_25, but can slightly shift with respect to it.
This means that the value of $f_{\rm cover}$ for \emph{IPTA}\_30
cannot be predicted from the respective value in \emph{IPTA}\_25.
This effect can also be seen in the second and third panels of
Figure \ref{fig:resultscrossmatches}, where we examine the
entire sample. 

More specifically, Figure \ref{fig:resultscrossmatches} summarizes
separately for each of the 30 well-localised binaries the evolution
of the main results (localisation area, coverage fraction
$f_{\rm cover}$, and number of potential hosts both for ETGs and
AGN) between the two IPTA configurations. We show the results
obtained with the \emph{IPTA}\_25 configuration with round
markers, and the ones obtained with \emph{IPTA}\_30 with squares.
The colour of each marker denotes the S/N obtained with \emph{IPTA}\_25,
and the sources are sorted horizontally as a function of their total
mass, in ascending order.

\begin{figure*}
    \centering
    \includegraphics[trim= 0 0 0 0
    ,clip,width=0.98\textwidth]{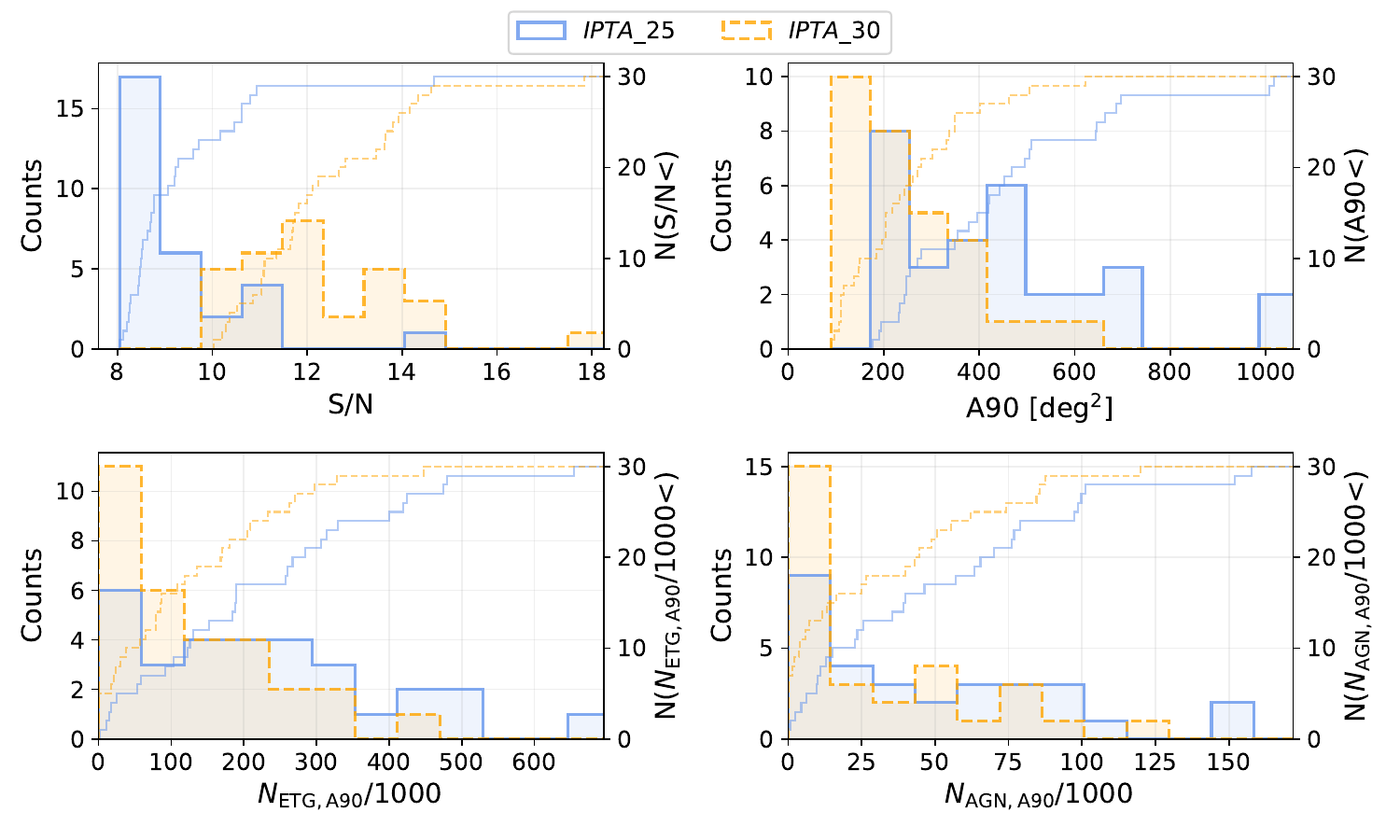}
    \caption{Distributions of the S/N (upper left panel), size of
    A90 (upper right panel), number of potential ETG hosts (lower
    left panel), and AGN hosts (lower right panel) for the 30
    well-localised binaries. The solid lines correspond to
    \emph{IPTA}\_25, while the dashed ones to \emph{IPTA}\_30.
    For both configurations we show the counts per bin on the
    left-hand side of the vertical axis, and the non-normalized
    cumulative distribution function on its right-hand side.}
    \label{fig:hists_combined}
\end{figure*}

Among the 30 well-localised binaries, the average increase in S/N
between \emph{IPTA}\_25 and \emph{IPTA}\_30 is approximately 32
percent. In the upper-left panel of Figure \ref{fig:hists_combined},
we show the distributions of S/N for both configurations. In every
panel of the figure we show both the counts per bin for the correspondent
quantity, and its non-normalized cumulative distribution function (the
values of which are reported on the right-hand side of the vertical axis).
The solid blue lines show the results obtained with \emph{IPTA}\_25,
while the dashed orange lines correspond to \emph{IPTA}\_30. 

The size of A90 constrained by \emph{IPTA}\_30 is on average $\approx38$
percent smaller than the one constrained by \emph{IPTA}\_25. This is in
line with the increase in S/N, since the localisation area is expected
to be approximately inversely proportional to the square of S/N, i.e.
A90 $\propto{\rm S/N}^{-2}$ \citep{Sesana10}. Therefore, an increase
in S/N of $\approx 30$ percent is expected to cause a decrease of
$\approx 40$ percent in the size of A90. The quartiles of the distributions
of A90 constrained with \emph{IPTA}\_30 for the 30 well-localised binaries
are ${\rm A90}=212_{140}^{325}$ ${\rm deg}^2$. The distribution of the size
of A90 is shown both for \emph{IPTA}\_30 and for our fiducial configuration
in the upper right panel in Figure \ref{fig:hists_combined}. All 30
binaries that are well-localised in our fiducial configuration have a size
of A90 smaller than 1,000 ${\rm deg}^2$ in \emph{IPTA}\_30. We note that
this is also true for the 8 binaries that, while having a $S/N\geq8$,
had a size of A90 larger than 2,000 ${\rm deg}^2$ in \emph{IPTA}\_25.
It is also worth noting that for three binaries the constraints on the
localisation area remain almost unchanged (see the upper panel of Figure
\ref{fig:resultscrossmatches}), when adding the extra 42 pulsars and 5 years 
of monitoring, while for one case, they get worse. While it is not trivial
to identify the exact cause of this, we hypothesize that this is likely due
to the stochasticity involved in the generation of the TOAs (where we
create random realizations of the pulsar noises, of the measurement white
noise, and of the GW stochastic background), and more importantly in the
MCMC sampling of the posterior distributions of the binary parameters. 

Regarding the coverage factor, we find that in 13 and 11 cases, A90 is
entirely within the footprint of the ETG and AGN catalogue, respectively.
These include all 9 that had this property in \emph{IPTA}\_25. However,
there are 5 and 4 unfortunate cases, where the new localisation area
ends up being completely within the sky region not covered by the ETG
and AGN catalogues, respectively.

We cross-match the sky maps constrained by \emph{IPTA}\_30 with
the ETG and AGN catalogues. The three quartiles of the distribution
of the number of ETGs within A90 are
$N_{\rm ETG}^{IPTA\_30}=85,622_{26,228}^{200,018}$. The same values
for the AGN scenario are $N_{\rm AGN}^{IPTA\_30}=14,793_{1,660}^{50,302}$.
These numbers also include the cases with $f_{\rm cover}=0$. We show
the distributions of the number of potential hosts both for
\emph{IPTA}\_25 and \emph{IPTA}\_30 in the lower left panel of Figure
\ref{fig:hists_combined} (for ETGs) and in the lower right panel (for
AGN). The presence of cases for which $f_{\rm cover}=0$ (and thus there
are zero cross-matched galaxies in A90) are marked by the fact that the
cumulative distributions on the number of potential hosts for
\emph{IPTA}\_30 start with a non-null value on the vertical axis.

Next, we focus on the subset of PTA sources the localisation area of
which is fully covered by EM surveys (i.e. $f_{\rm cover}=1$) in both
\emph{IPTA}\_25 and \emph{IPTA}\_30 to see how the number of potential
hosts evolves. This selection based on $f_{\rm cover}$ eliminates
biases due to incomplete coverage. The number of hosts within A90 is
decreased by 37.7 percent for ETGs and by 37.1 for AGN. This exactly
follows the fractional decrease in the size of A90 for the same binaries
($\approx 37.8$). 

Finally, with respect to the constraints of the remaining binary
parameters in \emph{IPTA}\_30, we obtain an average interquartile
range in logarithmic space for the frequency distribution of 0.01,
for the strain distribution of 0.17, and for the chirp mass distribution
of 1.10. As expected, the posterior distributions on these parameters
get in general narrower when we add the extra 42 pulsars and 5 more
years of observations, but the uncertainty on the chirp mass decreases
on average only by a factor of $\approx 9$ percent. Therefore, even
in \emph{IPTA}\_30, the chirp mass uncertainty remains significant,
and drives the uncertainty on the derived properties of the hosts.

\subsection{Missing galaxies}
\label{sec:missing_gals}
The EM catalogues we use to cross-match the GW localisation areas are
incomplete in sky regions with high contamination of stellar content
and/or dust, e.g., close to the Galactic plane. For each well-localised
binary, we estimate the number of potential hosts that are not present
in our EM catalogues. We do so by first calculating the number of pixels
in A90 that are not covered by EM surveys. Then we multiply this by
the average number of galaxies per pixel, calculated by dividing the
total number of entries in each catalogue by the number of pixels in
the respective catalogue footprint. 

\begin{figure}
    \centering
    \includegraphics[trim= 0 0 0 0
    ,clip,width=0.98\columnwidth]{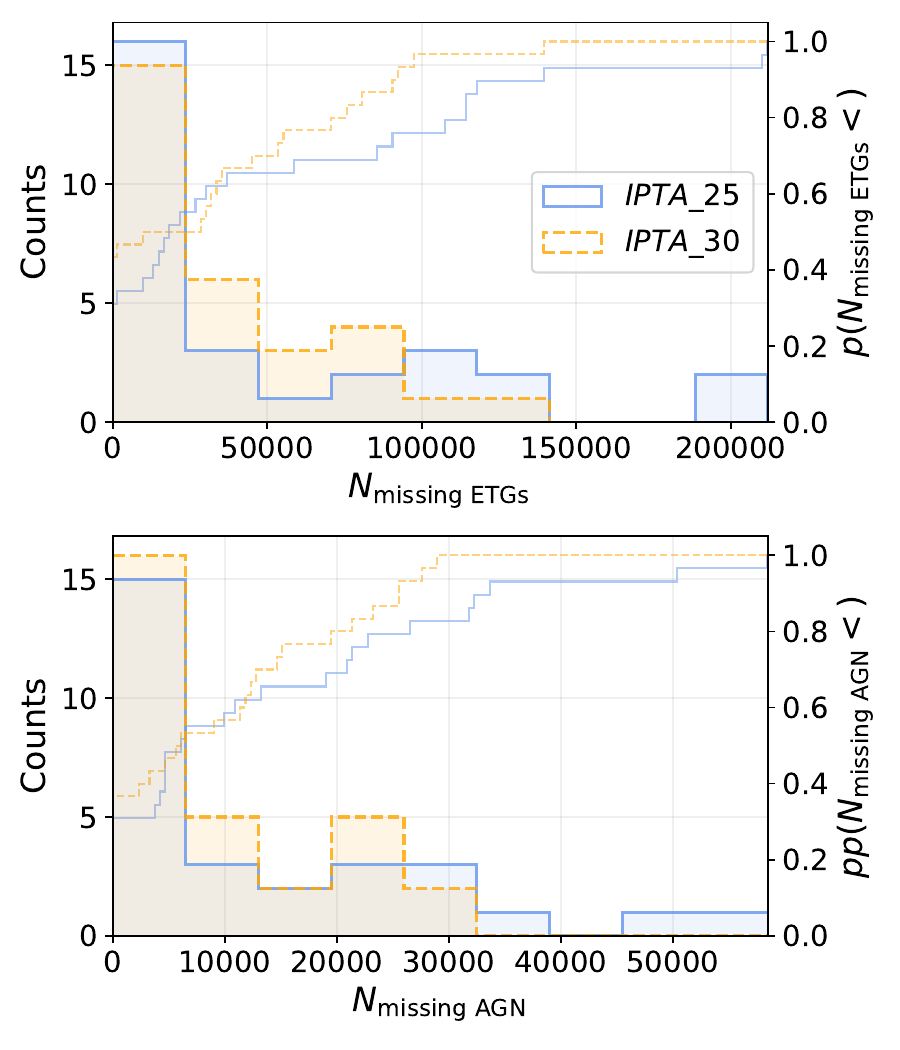}
    \caption{Distributions of the estimated number of potentially
    missing hosts due to the gaps in the EM catalogues. Blue solid
    lines show results obtained with \emph{IPTA}\_25, and orange
    dashed lines with \emph{IPTA}\_30. The upper panel shows the
    estimated number of missing ETGs, while the lower one the
    estimated number of missing AGN.
    \label{fig:missing_gals}}
\end{figure}

The distributions of the number of missing galaxies are shown in Figure
\ref{fig:missing_gals}. As before, the blue solid lines show the results
obtained with our fiducial PTA configuration, while the orange dashed
lines represent results obtained with \emph{IPTA}\_30.
The quartiles of the distributions of missing ETG hosts, calculated
including both cases with $f_{\rm cover}=1$ and with $f_{\rm cover}=0$,
are $N_{\rm ETG}=20,196_{0}^{103,506}$ for \emph{IPTA}\_25 and
$N_{\rm ETG}=19,355_{0}^{55,118}$ for \emph{IPTA}\_30. The same
estimates for the AGN case are $N_{\rm AGN}=6,421_{0}^{22,475}$ and
$N_{\rm AGN}=5,946_{0}^{15,102}$. The null values of the 25$-th$
percentiles are caused by the sources whose GW maps are fully covered
by the EM catalogues ($f_{\rm cover}=1$), and thus are associated to
zero missing galaxies based on our definition \footnote{Some potential
hosts might still be not present in the catalogues we use due to high
level of obscuration or flux limits of the surveys the catalogues we use
are obtained from.}. As can be seen in Figure \ref{fig:missing_gals},
the variance of these values is large, as it depends both on the size
of A90 and on the value of $f_{\rm cover}$ for each binary. 

The gaps in EM coverage highlighted by the large number of missing
galaxies can present a significant limitation in future searches for
the host galaxy. The real host may coincide with one of the missing
galaxies, making a multi-messenger detection impossible without
follow-up observations to cover these gaps. This effect also needs to
be accounted for both in the estimation of false-positive probabilities
for the binary-host association, and in any potential ranking systems
which associates to each candidate an absolute probability of being the
true host.

\subsection{Host ranking}
\label{sec:resranking}
Finally, for each galaxy within the localisation area we calculate the
host score. We use this to derive the relative ranking of ETGs and AGN,
independently for the two sets of galaxies. 

First, we identify galaxies that have $p_{cat}=0$ and thus can be excluded
as potential hosts, since their EM properties are inconsistent with the
GW posteriors. On average, 47.21 percent of the ETGs are associated to a
null score with \emph{IPTA}\_25, and this fraction is further increased to
61.14 percent with \emph{IPTA}\_30. The quartiles of the distribution of
the number of remaining ETGs within A90 are
$N_{\rm ETG}^{IPTA\_25}=86,587_{36,315}^{146,238}$ for \emph{IPTA}\_25
and $N_{\rm ETG}^{IPTA\_30}=23,950_{3,706}^{68,837}$ for \emph{IPTA}\_30.
On the other hand, the fraction of AGN that can be excluded based on their
EM properties is significantly lower, 0.26 percent and 3.76 percent for
\emph{IPTA}\_25 and \emph{IPTA}\_30, respectively. Thus the same quartiles
for the remaining AGN are $N_{\rm AGN}^{IPTA\_25}=39,655_{11,786}^{76,932}$
and $N_{\rm AGN}^{IPTA\_30}=12,828_{1,660}^{48,692}$.

The significant difference in our ability to exclude different types of
hosts based on their EM properties is caused by the different information
provided in the two employed catalogues. In particular, a null score for an
ETG can be caused by a null value of $p_{\rm M_{tot}}$ or $p_{\rm d_{lum}}$,
caused by a high discrepancy between the estimates of the luminosity distance
and, even more importantly, of the total mass obtained with the GW analysis
and the respective values in the EM catalogue. For instance, in the ETG
catalogue, a high fraction ($\approx$83 percent) of the galaxies have an
estimated SMBH mass smaller than $10^{8.5}M_\odot$, lower than any injected
binary mass. For this illustrative example we choose $10^{8.5}M_\odot$ (even
though the lowest simulated mass is higher than $10^{9}M_\odot$) because the
posteriors on total mass are very wide, and we only get a null value of
$p_{\rm M_{tot}}$ if the galaxy has an estimated value significantly
incompatible from such distribution. In the AGN ranking, the information
coming from the total mass and the luminosity distance is combined in the
calculation of $p_{m_{\rm W1,AGN}}$, which compares the apparent magnitudes
of the hosts estimated from the GW analysis and those listed in the catalogue.
This degeneracy, along with the poorly constrained chirp mass from the GW
pipeline leads to very wide distributions of apparent magnitudes, resulting
in a much lower fraction of null  $\mathcal{H}_{\rm AGN}$ host scores.

Hereafter, we show results of the host ranking obtained after all potential
hosts with a null score are discarded. First, we demonstrate the results of
the ranking method obtained using the same example source as in the previous
section. Then we show how these results are generalized for the entire
population of simulated and localised binaries.

\begin{figure}
    \centering
    \includegraphics[trim= 0 0 0 0
    ,clip,width=0.98\columnwidth]{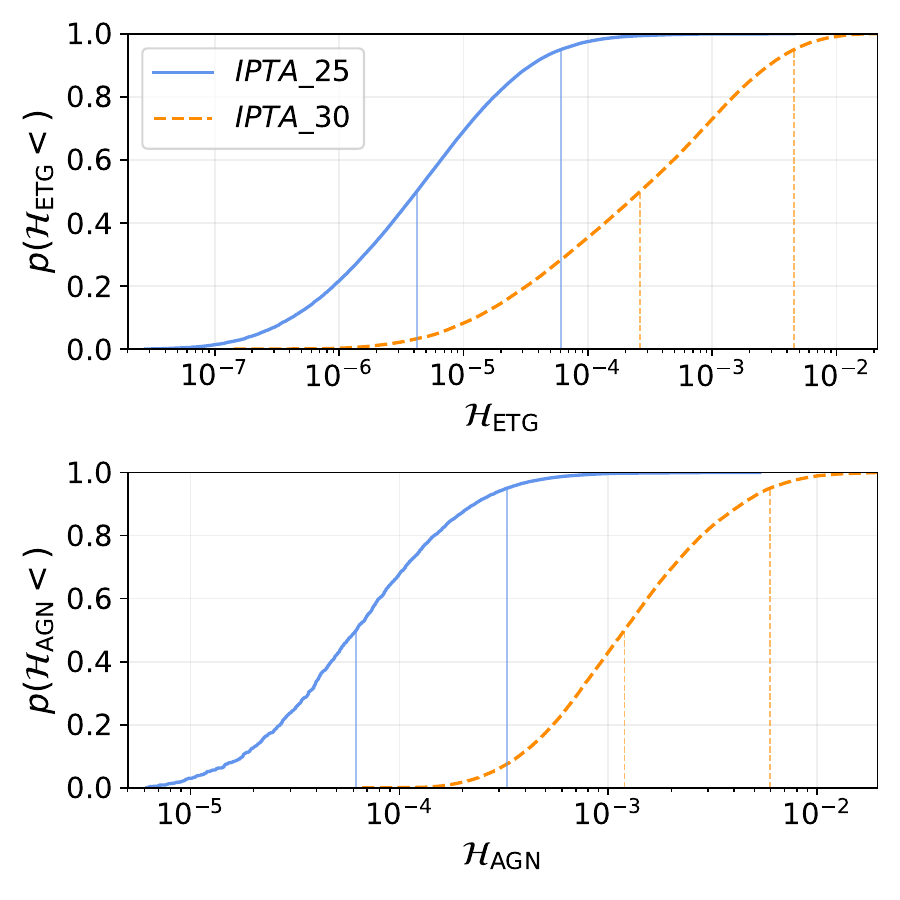}
    \caption{Cumulative distribution functions of the non-null
    host scores for the ETGs (upper panel) and AGN (lower panel)
    contained in A90 for the source used as an example in Section
    \ref{sec:constraintsimprov}, the position of which is shown
    in Figure \ref{fig:binposits380}. Blue solid (orange dashed)
    lines indicate the distributions for \emph{IPTA}\_25
    (\emph{IPTA}\_30).}
    \label{fig:scores380}
\end{figure}

After excluding galaxies with a null score, the localisation area of this
binary has a total of 206,543 ETGs with our fiducial configuration, which
is reduced to 190,176 for \emph{IPTA}\_30, significantly lower than the
initial number of ETGs within A90 (see Section \ref{sec:constraintsimprov}).
For our fiducial configuration, for example, this selection removes
approximately 51 percent of the galaxies in A90. The remaining AGN within
A90 are 90,132 for \emph{IPTA}\_25 and 62,203 for \emph{IPTA}\_30, only
slightly lower than the initial total number. Figure \ref{fig:scores380}
shows the cumulative distribution functions of the non-null host scores
for the example binary. The vertical lines highlight the medians and the
95$-th$ percentiles of the distributions. The efficiency parameter (i.e.
the ratio between the score of the top performing galaxies and the average
galaxies) for ETGs is $\mathcal{E}_{95-50}=14.44$, while for the AGN
scenario goes down to $\mathcal{E}_{95-50}=5.26$. For \emph{IPTA}\_30, we
see an improvement for ETGs, with an efficiency parameter
$\mathcal{E}_{95-50}=17.33$, but this improvement does not extend to AGN,
for which the efficiency parameter is $\mathcal{E}_{95-50}=4.98$. We remind
the reader that a higher value of this ratio indicates that the top galaxies
differ more from the lowest ranked and thus we may need to follow-up fewer
of these potential hosts to find the true one. However, since the calculated
scores do not correspond to an absolute probability of a galaxy being the
binary host, we cannot estimate how many potential hosts have to be followed
up to reach a certain likelihood for finding the host.
The aforementioned differences that allow us to exclude more ETGs
compared to AGN can also explain the higher efficiency in the ranking
of the two types of galaxies. We further discuss this in Section
\ref{sec:discus}.

\begin{figure}
    \centering
    \includegraphics[trim= 0 0 0 0
    ,clip,width=0.98\columnwidth]{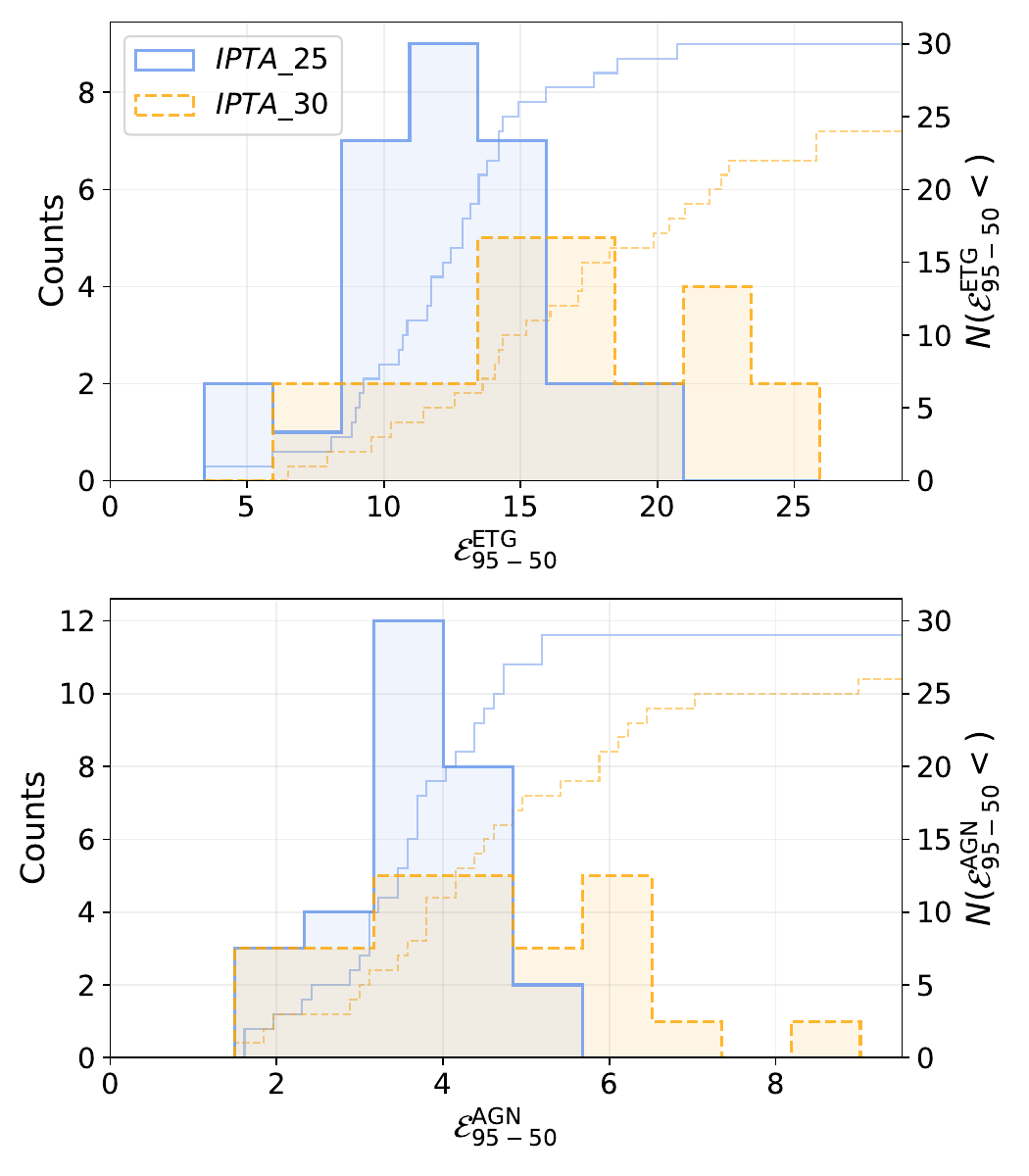}
    \caption{Distribution functions of the ranking efficiency
    $\mathcal{E}_{95-50}$ (i.e. the ratio between the 95-th and
    the 50-th percentiles of the non-null host score distributions)
    for the well-locsalized binaries with $f_{\rm cover}>0$. The
    plot on the top shows the results of the cross-matches between
    GW sky maps and the ETG catalogue, while the lower the results
    obtained from the cross-matches with the AGN catalogue. The blue
    solid line shows the results of fiducial PTA configuration,
    while the orange dashed lines the ones of \emph{IPTA}\_30.}
    \label{fig:scoreratios}
\end{figure}

The difference in the ranking efficiency between the two galaxy types
is consistently seen in the entire sample of well-localised PTA
sources. We then calculate the ratio $\mathcal{E}_{95-50}$ for all
the cases with $f_{\rm cover}>0$, and show its distribution for ETGs
(upper panel) and for AGN (lower panel) in Figure \ref{fig:scoreratios}.
In the ETG scenario the quartiles of the distribution of the efficiency
parameter are $\mathcal{E}_{95-50}^{\rm ETG}=12.41_{10.06}^{14.11}$ for
our fiducial PTA configuration and
$\mathcal{E}_{95-50}^{\rm ETG}=17.20_{13.65}^{21.06}$ for \emph{IPTA}\_30.
For AGN the same values are $\mathcal{E}_{95-50}^{\rm AGN}=3.65_{3.20}^{4.42}$
and $\mathcal{E}_{95-50}^{\rm AGN}=4.36_{3.51}^{5.81}$. We note that for
both types of potential hosts the ranking efficiency is improved (albeit
marginally for AGN) due to better constraints on the binary parameters
(especially the mass) obtained with \emph{IPTA}\_30. This is encouraging,
especially since the starting number of galaxies is already smaller due to
better localisation. This is further discussed in Section \ref{sec:discus}.

\section{Discussion}
\label{sec:discus}
In this section we discuss the implications of our findings and
highlight assumptions and caveats of the method we employ. We also
present some alternative approaches that might be explored in future
studies.

\subsection{Prospects for binary detectability in the near future}
\label{sec:discus:prospects}
We generate 1,000 binary populations, consistent with the GW background
measured in NG15 and select the loudest binaries requiring that there
was no evidence for their presence in NG15 (see Section \ref{sec:gwselec}).
We find that a significant fraction of the populations ($\approx$40
percent) could produce a detectable binary in the next 5 years. This
fraction increases to over 50 percent in \emph{IPTA}\_30. However, when
imposing the additional requirement of $S/N\geq8$ to be able to
localise the source, only 3.8 percent of the populations pass this cut
in \emph{IPTA}\_25. When simulating their GW signals and running the
detection analysis, this fraction is lowered to 3 percent, given that
some sources have an A90 greater than 2,000 ${\rm deg}^2$ .

Even though this may appear like a pessimistic prediction for the
prospects of multi-messenger discoveries in the near-future, it is
not entirely surprising. By the nature of the experiment, the S/N
of individual binaries, and thus our ability to constrain their sky
location, will increase gradually. Therefore, the most likely scenario
is that PTAs will first find evidence for the presence of a continuous
wave source in their data, and constrain its frequency and strain, but
will not be able to pinpoint its location initially \citep{Petrov25}.
However, with the addition of more pulsars and the extension of baselines
we will be able to localise the source with improving precision as more
and more data are collected. For instance, we see a $\approx40$ percent
decrease in the size of A90 between \emph{IPTA}\_25 and \emph{IPTA}\_30.
We also note that in this configuration, the loudest binary is potentially
localisable in approximately 15 percent of the realizations.

We also see that the predicted population of resolvable binaries in
\emph{IPTA}\_25 is clustered in the sky. This is clearly evident in
Figure \ref{fig:binposits}. This interesting finding suggests that
an upcoming detection would have a high probability of arising from
a specific region in the sky, more specifically the southern hemisphere.
This clustering is caused by our requirement that such a source was not
detectable by NG15. As mentioned above, this region experiences a
greater boost in sensitivity compared to NG15, because in \emph{IPTA}\_25
we include pulsars in the southern hemisphere from MPTA and PPTA.
The same clustering is present, but much weaker, in the binaries
potentially localisable with \emph{IPTA}\_30.

Since we have not imposed the non-detectability selection using only
the third data release of PPTA or the first data release of MPTA, we
cannot conclude whether our prediction regarding the observed clustering
is robust. However, it is safe to assume that had we imposed the same
criterion for the EPTA dataset, we would observe a similar trend, since 
there is significant overlap between the NANOGrav and EPTA pulsars. We
note, however, that the PPTA dataset includes fewer pulsars, while the
MPTA data have shorter baselines and thus may have less constraining
power. Therefore, it is possible that even if we performed similar cuts
on their detectability based on PPTA or MPTA, the remaining sources
could still end up mostly in the southern hemisphere (even though some
of the 38 sources would be excluded).

We emphasize that the main goal of the paper is not to predict the exact
properties of the first resolvable binary, but rather to quantify
their localisability and the number of viable host galaxies. A future
study, one specifically aimed to predict the parameters of the first
source that will be detected above the background, should impose as
criterion that these sources were not detected by any given PTA. It
should also be constrained by the findings of the search in the upcoming
third data release of IPTA.

However, if this clustering feature persists, we need to consider the
limitations imposed by the overlap of this region with the Galactic
plane, where the search for binary hosts is heavily limited by gaps
in the coverage of EM surveys. Even so, we find encouraging that the
localisation area of more than 25 percent of the well-localised binaries
is fully covered by both catalogues (see Figure \ref{fig:resultscrossmatches}).

Finally, we note that in some simulated SMBH binary populations more
than one binary is considered either detectable or even potentially
localisable according to our selection criteria. Since the focus of
this work is to quantify the number of potential hosts of the first
resolved PTA source and then rank them, for each binary population
we select only the top binary (in terms of S/N) to simulate its TOAs
and perform a detection analysis. However, since a number of them
have one or more "loud siblings", we highlight the need to further
implement detection pipelines that simultaneously search for multiple
binaries on top of the stochastic background \citep{Petiteau13,Curylo26}.

\subsection{localisation areas and cross-matches}
We perform realistic simulations of continuous-wave sources on top
of the GW background and constrain their parameters with a standard
GW detection pipeline. We find the median of A90 for the 30 sources
localised by \emph{IPTA}\_25 to be ${\rm A90}=408\ {\rm deg}^2$,
reduced to ${\rm A90}=212\ {\rm deg}^2$ in \emph{IPTA}\_30, as can
be seen from the upper right panel of Figure \ref{fig:hists_combined}.
Even though the localisation area decreases by $\approx40$ percent on
average, it remains large even in the extended PTA configuration.
We note that the poor localisation is typical for GW experiments across
the entire spectrum. These constraints on A90 are consistent with
findings in previous studies (e.g., \citealt{Goldstein19}
\citealt{Petrov24}, \citealt{Gardiner25}, and \citealt{Truant26}),
and inevitably lead to a large number of candidate hosts, i.e. hundreds
of thousands of ETGs and tens of thousands of AGN. Specifically, for
the 30 well-localised binaries the median number is 189,286 for ETGs
and 39,655 for AGN.

These numbers, which are different with respect to the ones listed
in previous studies \citep[e.g.,][]{Goldstein19, Petrov24}, derive
from the choice of all-sky surveys we use for the cross-match.
The NANOGrav catalogue presented in \citep{Arzoumanian21} and used in
\citet{Petrov24}, for example, contains a total of 43,532 galaxies
within 500 Mpc from Earth. The catalogue of ETGs we use has a factor
of approximately 450 more entries. While not complete over the entire
sky, the EM catalogues we adopt have a remarkable depth and uniformity
outside the Galactic plane, and therefore reflect a realistic scenario
for future multi-messenger searches. The main difference in the galaxy
catalogue used to test the host ranking method first presented in
\citet{Goldstein19} is that we use observed objects, while in the
aforementioned work the candidates are extracted from the Millennium
Run cosmological simulation \citep{Springel05}, and have a stellar mass
which is known \emph{a priori} and is not derived from observed magnitudes.

The remarkable number of candidate hosts within every localisation
area, even after having discarded the ones with a null host score,
highlights the need for an accurate estimation of the false-positive
probability for any future search for an EM counterpart. The source-host
association will rely on a specific EM signature (or a combination of
many), and its false-positive probability will depend on several factors,
including the involved observables, and the quality of the available EM
data. Many of the proposed binary signatures in AGN (e.g., periodic
variability or broad line Doppler shifts) can be confused with the
typical AGN behavior \citep{Vaughan16,Krishnan21,DOrazio23}. In
\citet{Davis24}, for example, is shown that the most luminous single
quasars can efficiently produce binary-like signatures in simulated
light-curves of the Rubin Observatory Legacy Survey of Space and Time
(LSST, \citealt{Ivezic19}), and this can lead to false-positive rates
as high as 60 percent for such objects. Similar issues may arise for ETG
hosts, which might contain a PTA source without exhibiting any distinct
observable EM property. For instance, the presence of a binary can be
revealed by detailed observations of morphological and kinematic
properties of the galaxies \citep{Bardati24}, while in typical images
of average quality they may look indistinguishable from the rest of the
ETGs. The information coming from any of these observables can be
introduced in the $p_{\rm cat}$ term of the host score, making the
ranking system more efficient, further reducing the number of potential
hosts to follow up.

We also note that the localisation areas and the resulting cross-matches
with the EM catalogues are obtained using sky maps with the resolution
of a Healpix map with NSide=32. This choice, which likely has negligible
impact on our main results, was made to avoid holes in the GW maps (i.e.
pixels excluded from A90, while fully surrounded by pixels that were
included) caused by a limited amount of posterior samples. Increasing the
resolution to NSide=64 would lead to localisation areas without a compact
shape in several cases. This feature is also present in some A90 calculated
with NSide=32, but negligibly, so it does not affect the cross-matches or
the ranking. Future studies could use longer MCMCs and opt for a higher
value of NSide, with the increased resolution likely resulting in slightly
smaller A90s.

Finally, the precision of PTA sources localisation, together with the one
of the chirp mass constraints, could be greatly improved by fully leveraging
the information contained in the pulsar term of the GW signal, which would
be possible with accurate pulsar distances measurements to use as priors,
and detection pipelines that don't consider the phase of such term as nuisance
parameter, but rather derive it from the inferred pulsar distances themselves
\citep{Taylor26}. This is particularly true for systems with high masses and
low luminosity distance, which are loud enough to be soon detected by PTAs
and have an orbital evolution which happens on shorter timescales with respect
to the rest of the population \citep[see, e.g., ][]{Mcgrath:2022,DengFinn:2011}.
This works therefore highlights the importance of better pulsar distances
constraints and of the development of efficient "phase-linked" pipelines for
the future of multi-messenger analyses of PTA sources.

\subsection{Efficiency of host ranking}
The method we develop allows us to rank the candidate hosts based on
their sky position and their EM properties. Using additional information
on top of their sky positions can facilitate future multi-messenger
searches for the host galaxy of the PTA source. In fact, we can exclude
approximately half of the ETGs within A90 (the ones with null scores)
based only on their photometric properties.

Figure \ref{fig:scoreratios} shows that the top performing ETGs in
our ranking can be tens of times more likely to contain the PTA source
than the typical galaxies within A90. In the ETG case, the total mass,
$M_{\rm tot}$, and the luminosity distance, $d_{\rm lum}$, enter
separately the calculation of $\mathcal{H}_{\rm ETG}$ (see Equation
\ref{eq:scoreetg}), thus providing more constraining power. In contrast,
the best performing AGN have a score approximately 4 times higher than
the median. This is approximately the ratio we would obtain considering
only the sky positions; the additional constraints from the apparent W1
magnitude do not significantly improve the ranking. This is because
$m_{\rm W1, AGN}$ is an observable which combines the total mass of the
binary and its luminosity distance from Earth (which are not listed in
the AGN cataloguee we adopt), resulting in very broad posteriors.
Incorporating other sources of uncertainty (e.g., the Eddington fraction,
the bolometric correction) makes this property even less constraining.
Our inability to efficiently rank AGN may not be a huge limitation,
since previous studies suggested that detectable AGN are less likely
to host PTA sources, compared to quiescent massive galaxies \citep{Truant26}.
However, in this work we conduct the whole analysis for both types of
hosts in parallel for completeness. 

Based on the above, we conclude that the ranking method can be boosted
with a few improvements in the EM catalogues. For instance, building an
AGN catalogue that provides estimates of the redshift and the mass of the
central SMBH for each galaxy would result in a ranking efficiency similar
to the ETG case. At the same time, the ranking efficiency for ETGs could
be further improved if the catalogue provided the galaxy morphological
type. We note that the ETG catalogue excludes only AGN and contains all
types of quiescent galaxies. We made an implicit assumption that all of
those are ETGs, by assigning the entirety of the stellar mass to the
bulge when calculating the mass of the central SMBH (see Section
\ref{sec:etgcat}). This assumption does not hold for spiral galaxies,
in which only a fraction of the stellar component is in the bulge. Late
type galaxies are present in the catalogue, although they are typically
on the lower end of the stellar mass distribution, and the majority may
have been excluded as potential hosts with null scores. However, it is
possible that some remain even after this cut, and thus our estimates on
the total number of potential hosts can be considered as upper limits.
Practically, a smaller bulge mass would translate into a smaller SMBH
mass, and could be more likely rejected as a potential host due to the
incompatibility with the (typically high) total binary mass estimates
obtained from the GW analysis. 

We emphasize that the main limiting factor comes from the GW analysis
and the wide posteriors of the chirp mass \citep{Veronesi25}. Even
though these posteriors improve in \emph{IPTA}\_30, they remain wide.
This large uncertainty propagates to the derived posteriors of total
mass, luminosity distance, and W1 apparent magnitude (for the AGN case).
As a result, in some cases the sky position almost exclusively determines
the ranking. However, as we have already seen, for a significant fraction
of candidate hosts, including the information encoded in $p_{\rm cat}$
is crucial, since it can reject a considerable fraction of galaxies and
contribute to the relative ranking of the remaining candidates. 

Finally, as mentioned in Section \ref{sec:ranking}, the host scores
defined by Equations \ref{eq:scoreetg} and \ref{eq:scoreagn} only provide
the relative ranking of the hosts, and not the absolute probability that
a certain candidate contains the GW-detected source. In order to estimate
such a probability, one would need to take into account the normalization
of the various probability densities in the score, the total probability
density associated with the part of GW sky map that falls in the gaps of
the EM catalogues, and the completeness of the latter. The completeness can
be estimated as a function of luminosity and redshift by comparing the
number density of observed galaxies with the estimates from luminosity
functions (see, e.g. \citealt{Veronesi25b}). This is non-trivial and thus
the calculation of an absolute probability for each potential host is
beyond the scope of this work, and will be presented in a follow-up
publication \citep{Petrov26}.

%------------------------------------------------------------------
%------------------------------------------------------------------
%------------------------------------------------------------------

\section{Summary and Conclusion}
\label{sec:concl}
In this work, we perform realistic simulations to predict the properties
of the first PTA source that can be resolved on top of the GW background
in the near future. We focus on sources that can be localised with
sufficient accuracy to allow for multi-messenger searches. We cross-match
the GW maps with EM catalogues to estimate the number of potential hosts,
exclude galaxies based on their EM properties, and rank the remaining.

Specifically, we generate 1,000 realistic SMBH binary populations that
are compatible with the measured stochastic GW background in NG15 using
semi-analytical models from \texttt{holodeck}. We select the loudest
binary for each population and simulate their TOAs on top of the observed
background in realistic PTA configurations that resemble our expectations
for upcoming IPTA datasets. In particular, we consider three PTAs:
\emph{IPTA}\_20, \emph{IPTA}\_25, and \emph{IPTA}\_30, with baselines
of 20, 25 and 30 years, containing a total of 116, 158 and 200 pulsars,
respectively. We then use \texttt{QuickCW}, one of the standard GW
detection pipelines, to constrain the main properties of the injected
binaries. 

We cross-match the resulting sky maps with two all-sky catalogues of
ETGs and AGN, which are expected to be complete for the massive hosts
of PTA sources. We estimate the number of potential hosts for each type
of galaxy independently and, for the first time, quantify the impact
of incomplete EM coverage by estimating the number of missing hosts. We
convert the GW posteriors of binary parameters into EM observables that
match the available information in each galaxy catalogue. We then use this
information to discard potential hosts that are completely incompatible
with the expectations coming from the GW analyses. We rank the remaining
galaxies based on the probability of containing the PTA source. To do so,
we develop a ranking system that takes into account both the photometric
and the astrometric properties of each ETG and AGN.

Our main conclusions can be summarized as follows:
\begin{itemize}
    \item A total of 212 and 378 populations have a detectable binary
    ($S/N\geq4$) in \emph{IPTA}\_20 and \emph{IPTA}\_25, respectively.
    This number is increased to 507 in \emph{IPTA}\_30. Our estimated
    probability to resolve one of these detectable sources in the next
    ten years is therefore approximately 50 percent. These detectable
    sources are isotropically distributed in the sky with their main
    parameters (total mass, mass ratio, luminosity distance from Earth)
    following the underlying population of loudest binaries from each
    simulated population. Approximately 70 percent of the detectable
    binaries have a frequency $f\lesssim10$ nHz.
    \item A total of 3, 38, and 141 binary populations have a binary that
    reaches $S/N\geq8$ and is potentially localisable by \emph{IPTA}\_20,
    \emph{IPTA}\_25 and \emph{IPTA}\_30, respectively. Similarly to the
    detectable binaries, these sources typically have low frequencies. They
    also have lower luminosity distances from Earth and higher total masses
    with respect to the rest of the population. For \emph{IPTA}\_25 (our
    fiducial configuration), most of these localisable sources are clustered
    in the area with $90^\circ<$ RA $ <225^\circ$ and $-45^\circ<$ Dec $<0^\circ$.
    An over-density in the same sky region is also observed in \emph{IPTA}\_30,
    but the clustering is weaker.
    \item For our fiducial configuration, 30 binaries have a localisation area
    smaller than A90 = 2,000 ${\rm deg}^2$, and are considered well-localised.
    Their typical size of A90 spans hundreds of square degrees with a median
    of A90 = 408 ${\rm deg}^2$, containing hundreds of thousands of ETGs and tens
    of thousands of AGN. The median number of potential hosts within A90 is
    $\approx$190,000 for ETGs and $\approx$40,000 for AGN. The size of A90 and,
    in turn, the number of potential hosts is approximately 40 percent smaller
    in \emph{IPTA}\_30.
    \item In our fiducial PTA configuration 22 (21) of the 30 well-localised
    binaries have an A90 partially outside of the footprint of the ETG (AGN)
    catalogue. In \emph{IPTA}\_30 this slightly decreases with 13 for ETGs and
    11 for the AGN case having total coverage, while we also find cases (5 for
    ETGs and 4 for AGN) in which A90 is completely outside of the footprint of
    the surveys. Incomplete EM coverage implies that the true binary host might
    not be present in the EM catalogues. The median of the number of potentially
    missing hosts is $\approx$20,000 for ETGs and $\approx$6,000 for AGN in our
    fiducial configuration, and only slightly decreases in \emph{IPTA}\_30
    ($\approx$19,500 and $\approx$6,000).
    \item Our ranking method assigns a null score on average to
    approximately 47 and 61 percent of the ETGs for \emph{IPTA}\_25 and
    \emph{IPTA}\_30, respectively. These are galaxies with an estimated
    luminosity distance or central SMBH binary mass incompatible with the
    estimates obtained from the GW analysis. The selection based on the
    apparent W1 magnitude for AGN can reject only approximately 0.3 and 4
    percent of the potential hosts in \emph{IPTA}\_25 and \emph{IPTA}\_30,
    respectively.
    \item With our fiducial configuration, among the non-rejected hosts,
    the top performing ETGs (AGN) have a score which is $\approx12.4$
    ($\approx3.7$) times higher than the average candidate. A marginally
    higher ranking efficiency is achieved with \emph{IPTA}\_30, where the
    respective median values are $\approx17.2$ ($\approx4.4$). The lower
    ranking efficiency for AGN is due to the use of apparent magnitudes
    instead of separately using the total SMBH mass and the luminosity
    distance, like for ETGs. 
\end{itemize}

Thanks to the always improving PTA sensitivity, the first resolved source
has a non-negligible chance to be detected in the next 10 years. Once the
binary is localised, a ranking system like the one proposed in this work
will be an essential starting point for the binary host search. Our method
identifies which galaxies are the most promising and thus worth prioritizing
in searches for EM binary signatures. This work also highlights the need for
EM catalogues with sky coverages as complete as possible. The comparison of
results between the two catalogues shows that having estimates of the total
mass of the SMBH as well as of their redshift can boost the ranking efficiency
and exclude more candidate hosts. Therefore, building an AGN catalogue with
these quantities is important. Our chances of discovering the first
multi-messenger SMBH binary greatly depend on both the sensitivity of PTAs
and the completeness and uniformity of the available EM data. Since a perfectly
complete census of ETGs and AGN is impossible to obtain in the foreseeable
future, a detailed estimation of the catalogues' completeness (as a function
of luminosity, redshift, and sky position) will be crucial for correctly
estimating the false-positive probability of any future host-binary association.

%------------------------------------------------------------------
%------------------------------------------------------------------
%------------------------------------------------------------------

\vspace{-0.8em}
\section*{Acknowledgements}
The authors thank Emiko Gardiner and Bence Bécsy for their help
and support with the creation of binary populations and the
trouble-shooting with the MCMCs algorithms. MC acknowledges support
by the European Union (ERC, MMMonsters, 101117624). This research
used resources of the Center for Institutional Research Computing
at Washington State University. SRT is a member of the NANOGrav
collaboration, which receives support from NSF Physics Frontiers
Center award number 1430284 and 2020265. SRT acknowledges support
from an NSF CAREER \#2146016, NSF AST-2307719, NSF NRT-2125764, and
NASA LPS-80NSSC26K0342. SRT also acknowledges support from a Chancellor's
Faculty Fellowship from Vanderbilt University. DJD acknowledges support from Sapere Aude Starting grant No. 121587 through the Danish Independent Research Fund.
{\em Softwares}: 
\texttt{Numpy} \citep{Harris20}; 
\texttt{Matplotlib} \citep{Hunter07}; 
\texttt{SciPy} \citep{Virtanen20};
\texttt{Astropy} \citep{Astropy18};
\texttt{Healpy} \citep{Zonca19};
\texttt{Enterprise} \citep{Ellis20};
\texttt{QuicCW} \citep{Becsy22}.

%------------------------------------------------------------------
%------------------------------------------------------------------
%------------------------------------------------------------------
\vspace{-1.0em}
\bibliographystyle{mnras}
\bibliography{bibliography}

%------------------------------------------------------------------
% Don't change these lines
\bsp
\label{lastpage}
\end{document}